\documentclass[a4paper,fleqn]{cas-dc}

\usepackage{graphicx}
\usepackage[font=scriptsize, justification=centering]{caption}
\usepackage{subcaption}
\usepackage[linesnumbered, ruled,vlined]{algorithm2e}
\usepackage[switch]{lineno}
\usepackage{multirow}
\usepackage{enumitem}
\usepackage{listings}
\usepackage{xspace}
\usepackage{tcolorbox}
\usepackage{mdframed}
\usepackage[flushleft]{threeparttable}
\usepackage[numbers]{natbib}

\usepackage{color,soul}

\def\tsc#1{\csdef{#1}{\textsc{\lowercase{#1}}\xspace}}
\tsc{WGM}
\tsc{QE}
\tsc{EP}
\tsc{PMS}
\tsc{BEC}
\tsc{DE}

\newdefinition{definition}{Definition}

\definecolor{light-gray}{gray}{0.92} 
{\begin{mdframed}[backgroundcolor=light-gray,
skipabove=5pt,
skipbelow=0pt,
nobreak=false
]\begin{mdtheorem}{name}{label}}%
{\end{mdtheorem}\end{mdframed}}
\definecolor{ao}{rgb}{0.0, 0.5, 0.0}
\newcommand{\etal}{\textit{et al.}\space}
\newcommand{\tool}{\textsc{Openia}\xspace}
\newcommand{\magiccoder}{\textsc{Magicoder}\xspace}
\newcommand{\deepseek}{\textsc{DeepSeek Coder}\xspace}
\newcommand{\codellama}{\textsc{Code Llama}\xspace}
\newcommand{\gpt}{o4-mini\xspace}
\newcommand{\inhouse}{\textsc{In-house LLM-AJ}\xspace}
\newcommand{\external}{\textsc{External LLM-AJ}\xspace}
\newcommand{\llmcheck}{\textsc{LLM-Check}\xspace}

\begin{document}
\let\WriteBookmarks\relax
\def\floatpagepagefraction{1}
\def\textpagefraction{.001}

\shorttitle{\tool}

\shortauthors{Bui \textit{et~al.}}

\title [mode = title]{Correctness Assessment of Code Generated by Large Language Models Using Internal Representations}    

\author{Tuan-Dung Bui}
[orcid=0009-0007-7318-6896]
\ead{21020006@vnu.edu.vn}
\affiliation{organization={Faculty of Information Technology, VNU University of Engineering and Technology},
    city={Hanoi},
    country={Vietnam}}

\author{Thanh Trong Vu}
\ead{thanhvu@vnu.edu.vn}

\author{Thu-Trang Nguyen}
[orcid=0000-0002-3596-2352]
\ead{trang.nguyen@vnu.edu.vn}
\cormark[1]

\author{Son Nguyen}
[orcid=0000-0002-8970-9870]
\ead{sonnguyen@vnu.edu.vn}

\author{Hieu Dinh Vo}
[orcid=0000-0002-9407-1971]
\ead{hieuvd@vnu.edu.vn}

\cortext[cor1]{Corresponding author}

\begin{abstract}
Ensuring the correctness of code generated by Large Language Models (LLMs) presents a significant challenge in AI-driven software development. Existing methods predominantly rely on black-box (closed-box) approaches that evaluate correctness post-generation failing to utilize the rich insights embedded in the LLMs' internal states during code generation. 
This limitation leads to delayed error detection, increased debugging costs, and reduced reliability in deployed AI-assisted coding workflows.
In this paper, we introduce \tool, a novel white-box (open-box) framework that leverages these internal representations to assess the correctness of LLM-generated code. 
By systematically analyzing the intermediate states of representative open-source code LLMs, including DeepSeek-Coder, \codellama, and \magiccoder, across diverse code generation benchmarks, we found that these internal representations encode latent information, which strongly correlates with the correctness of the generated code.
%
%
Building on these insights, \tool uses a white-box/open-box approach to make informed predictions about code correctness, offering significant advantages in adaptability and robustness over traditional blackbox methods and zero-shot approaches. 
Our results show that \tool consistently outperforms baseline models, achieving higher accuracy, precision, recall, and F1-Scores with up to a 2X improvement in standalone code generation and a 3X enhancement in repository-specific scenarios. 
By unlocking the potential of in-process signals, \tool paves the way for more proactive and efficient quality assurance mechanisms in LLM-assisted code generation.
\end{abstract}



\maketitle

\section{Introduction}
The emergence of \textbf{L}arge \textbf{L}anguage \textbf{M}odels for Code (Code LLMs) has transformed the landscape of automated code generation, offering a promising solution to improve developer productivity and address the growing demand for software~\cite{survey-icse24,llm4code-survey,llm4code-survey2,llm4code-survey3}. 
These models, trained on vast repositories of code, are capable of automating repetitive tasks, suggesting optimizations, and even generating complex functionalities. 
As LLM-generated code becomes increasingly integrated into real-world systems, ensuring the quality and reliability of this code has become critical~\cite{msr2023-prem-llm-code-bugs,llm-code-bugs1,code-quality-chagpt,evaluating-chatgpt}. 
Low-quality code can lead to significant functional failures, severe security vulnerabilities, and increased maintenance costs, emphasizing the need for rigorous evaluation and quality assurance mechanisms~\cite{survey_3}.

However, recent studies show a concerning trend that LLM-generated code frequently contains more bugs and security vulnerabilities than human-written code~\cite{code-quality-chagpt, evaluating-chatgpt, security-issue-ccs2023,chatgpt-secure,code-llm-security,copilot-bugs,lost-at-c,msr2023-prem-llm-code-bugs}. 
This problem has led developers to find LLM-generated code often unusable, resulting in considerable time spent debugging or discarding it entirely~\cite{survey-icse24, expectation}.
To detect and mitigate these errors, current approaches rely on post-hoc methods, such as static or dynamic code analysis and testing, which can be resource-intensive and may miss context-specific issues.

{Unlike post-hoc approaches analyzing the final code only after code generation is \textit{complete}, an \textit{in-process} strategy takes advantage of the internal states capturing the ``thought process'' of code LLMs \textit{during} code generation, enabling early detection of potential issues.}
This is particularly promising for code LLMs, which offer a unique ``\textit{open-box}'' advantage to access code LLMs' internal states, especially with open-source code LLMs~\cite{deepseek-coder,codellama,magicoder}.
Unlike human coding, which remains a ``\textit{closed-box}'' process where only the final code is visible, and the developer's intermediate reasoning and thought process are inaccessible, code LLMs generate a wealth of internal representations and byproducts that can be examined in real-time. 
By leveraging these signals, we can move beyond passive validation and instead, proactively assess correctness as code is generated.
This opens an exciting opportunity: \textit{Can we leverage this visibility to proactively detect
during the code generation process, preventing bugs before they reach developers?}

While extensive research has been conducted on the performance and capabilities of code LLMs~\cite{code-quality-chagpt, evaluating-chatgpt,llm-evaluation,llm-evaluation2,chen2024evaluating,llm-code-quality}, there is a notable gap in understanding how their internal representations, specifically the hidden states during code generation, encode information related to reliability.
These internal states capture the model's intermediate understanding of the input and output, but their relationship with code quality remains largely unexplored. If these representations do indeed encode reliability signals, they could offer a novel pathway to assess and enhance code quality directly during generation rather than relying solely on external validation.

We hypothesize that \textit{the internal states of LLMs encapsulate meaningful signals that reflect the correctness of the code they generate}. These signals, embedded in the representations formed during the generation process, may correlate with key quality attributes.
Understanding and leveraging these signals could enable the proactive identification of faults, thereby improving the overall quality assurance pipeline.

In this work, we tackle the challenges and gaps in assessing the reliability of code generated by LLMs by presenting two key contributions.
\textbf{\textit{First}}, we introduce \tool, a novel framework that leverages the internal representations of code LLMs during code generation to assess the correctness of the generated code. Unlike traditional quality assurance techniques that rely heavily on post-hoc validation, \tool identifies and utilizes the signals embedded within the LLM's internal states. 
By doing so, it could enable earlier detection of potential bugs, reducing both the computational cost and the time required for quality assurance. 
%
%
\textbf{\textit{Second}}, we conduct a systematic analysis of the internal signals of code LLMs during code generation, focusing on three representative open-source models, including DeepSeek-Coder, \codellama, and \magiccoder. These models are evaluated across diverse benchmarks, encompassing standalone code generation and repository-level generation tasks. Our analysis reveals specific patterns and features in the internal representations that align with the reliability attributes of the generated code. 

Specifically, our experimental results demonstrate that \tool consistently outperforms traditional classification-based and zero-shot approaches across a range of benchmarks. For independent/standalone code generation tasks, \tool achieves up to a 2X improvement in accuracy compared to zero-shot baselines and surpasses classification-based methods, such as those utilizing CodeBERT and CodeT5+, by up to 20\% in accuracy. In repository-level code generation, \tool shows a remarkable, up to 3X, improvement in F1-Score, highlighting its robustness in handling complex and context-dependent coding scenarios. 
These findings provide valuable insights into how LLMs encode and process information related to code correctness, offering a deeper understanding of their inner workings.

By combining these contributions, this work lays the foundation for more efficient and proactive approaches to ensuring the quality of LLM-generated code. It highlights the latent capabilities of code LLMs to encode reliability-related information and demonstrates the potential of leveraging these signals to enhance quality assurance in automated software development.

\section{Internal Representations in LLMs for Code Generation}

Code LLMs rely on a sequence-to-sequence mapping mechanism, where input sequences (e.g., natural language descriptions or partially written code) are mapped to output sequences (e.g., complete or corrected code)~\cite{llm-survey, llm-evaluation,llm-evaluation2,llm}. Central to this process is the internal representations, which encode the intermediate understanding of the model as it processes inputs and generates outputs. These representations provide a window into the code LLM's decision-making process, offering insights into how the model predicts the next code tokens.

\subsection{The Code Generation Process}

For a code LLM $\mathscr{M}$, given an input sequence $x=\{x_1, x_2, ..., x_n\}$, where $x_i$ represents the $i$-th token in the input, $\mathscr{M}$ generates {a} code output sequence $c=\{c_1, c_2, ..., c_m\}$ token by token.
At each generation step $s$, the model predicts the next code token $c_s$ by leveraging both the input sequence and the tokens generated so far, denoted $c_{<s>} = \{c_1, c_2, ..., c_{s-1}\}$.
The prediction process at step $s$ can be represented as: 
$$P(c_s|x,c_{<s}) = \text{softmax}(W_o \cdot h_{l,s} + b_o)$$
where $h_{l,s}$ is the final hidden state of the LLM at layer $l$ for token $c_s$, $W_o$ and $b_o$ are the output layer weights and bias.

\subsection{Internal Representations}

\textit{Internal representations}, or \textit{intermediate activations}, are vectors $h_{l,s}$ that capture the latent information processed by the model at a specific layer $l$ for a given code token $c_s$. These activations are computed as:
$$h_{l,s} = f_l(h_{l-1,s}, \text{context}_s)$$
where $h_{l-1,s}$ is the activation from the preceding layer $(l-1)$. Meanwhile, $\text{context}_s$ represents contextual information (e.g., attention-weighted representations of other tokens), $f_l$ is the layer-specific transformation function (e.g., feedforward, attention, or normalization operations) applied at layer $l$.
Note that the initial hidden state for a token $c_s$ is $h_{0,s}$ is the embedding vector $e_s$ of $c_s$.

Each $h_{l,s}$ is a vector $\mathbb{R}^d$ where $d$ is the dimensionality of the LLM's hidden states (e.g., 4,096 in the case of \codellama). The sequence of activations across all layers and code tokens forms a hierarchical representation of the input and generated sequences that could encode syntactic, semantic, and contextual information at varying levels of abstraction.
Internal representations encapsulate the underlying structure and logic the model uses to generate code. By analyzing these representations, we gain insights into the model's ``thought process'' and ``intuition'' during code generation. These representations offer valuable cues about the correctness and reliability of the generated code $c$.


\section{A Framework Leveraging Internal Representations for LLM Generated Code Quality Assessment}

\subsection{Task Formulation}

Given a code LLM $\mathscr{M}$, an input prompt $x$, and the corresponding LLM-generated code $c$, the objective is to predict whether $c$ meets the correctness criteria as defined by the requirements provided in $x$. Specifically, the correctness of $c$ is evaluated by verifying whether $c$'s behaviors align with the intended behaviors specified in $x$.

Let $\mathcal{T} = \{t_1, t_2, \dots, t_n\}$ represent a set of validation tests designed to assess whether $c$ adheres to the requirements in $x$. Each test $t \in \mathcal{T}$ returns an outcome, denoted as $t(c)$, which indicates whether $c$ \textit{passed} or \textit{failed} the test. Code $c$ is considered correct if and only if it passes all the tests.
$$
\text{Correctness}(c) =
\begin{cases} 
1 & \text{if } \forall t \in \mathcal{T}, \; t(c) = \textit{passed}, \\ 
0 & \text{otherwise}.
\end{cases}
$$

The goal is to develop a predictive model $\mathscr{Q}$ that evaluates the correctness of generated code $c$ based on the requirements specified by the input prompt $x$ and the context provided by the code LLM $\mathscr{M}$. Formally:
$$
\mathscr{Q}(\mathscr{M}, x, c) = y
$$
where $y \in \{0, 1\}$ is the predicted correctness label for $c$, with $y = 1$ indicating that $c$ meets the correctness criteria and $y = 0$ otherwise.

\subsection{Framework Overview}

\begin{figure*}
    \centering
    \includegraphics[width=1.0\linewidth]{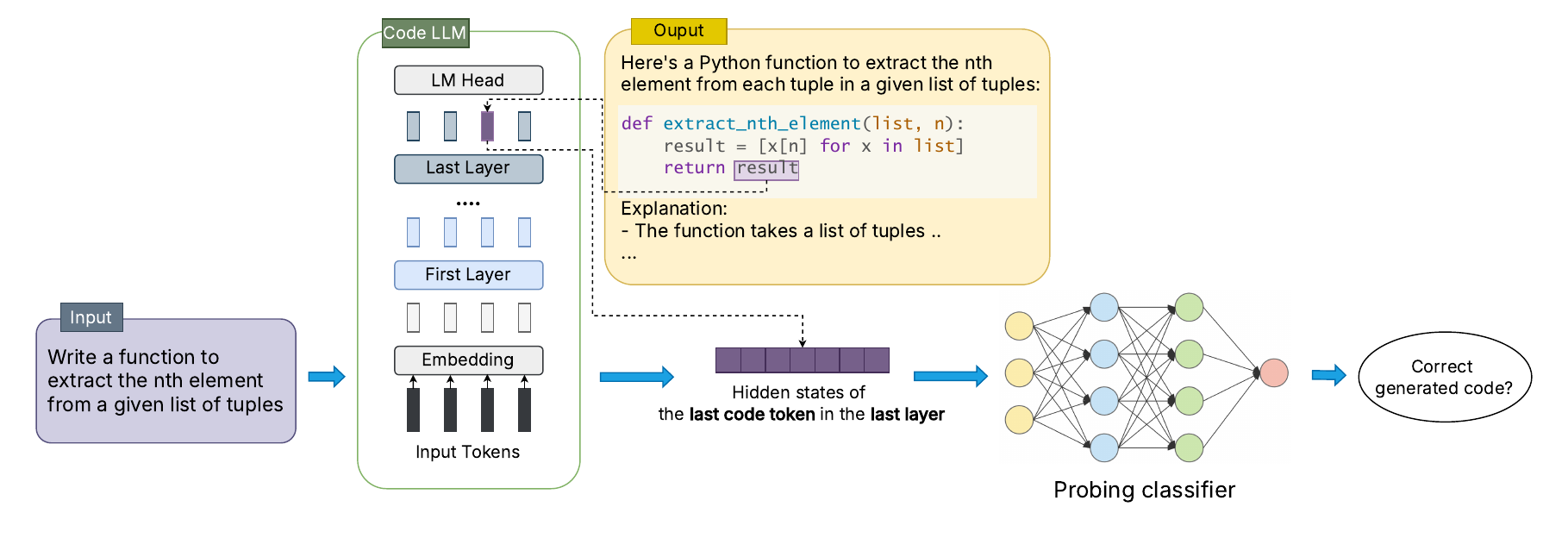}
    \caption{\tool's overview}
    \label{fig:framework_overview}
\end{figure*}

%

Fig.~\ref{fig:framework_overview} illustrates the overview of our framework, which leverages the internal representations of code LLMs during code generation to assess the correctness of its generated code. 
Specifically, given a code LLM $\mathscr{M}$ and its output code sequence $c=\{c_1, c_2, \dots, c_m\}$, \tool extracts the all internal states produced during the decoding process of $\mathscr{M}$: 
$$H = \big\{h_{l,s} | l \in \{1, ..., L\}, s \in \{1, ..., m\} \big\}$$
where $h_{l,s}$ represents the hidden state at layer $l$ corresponding to token $c_s$.
These internal representations are then fed to a probing classifier~\cite{probing}, which determines whether the generated code $c$ is correct.

\begin{figure}
\centering
\begin{subfigure}{0.48\textwidth}
\centering
 \includegraphics[width=1\columnwidth]{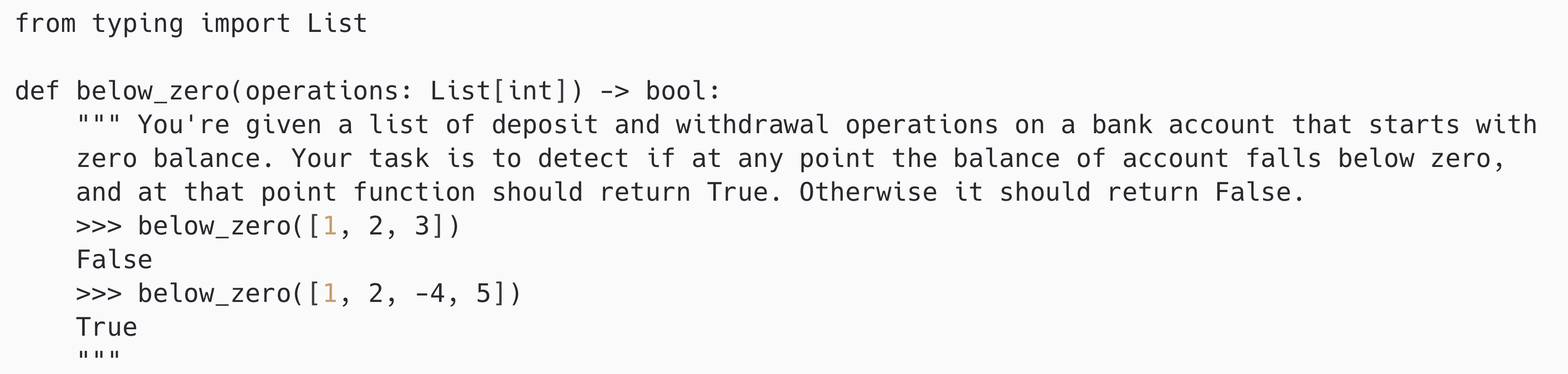}
\caption{Prompt provided to the model}
\label{fig:ex_standalone_input}
\end{subfigure}\\
\begin{subfigure}{0.48\textwidth}
\centering
 \includegraphics[width=1\columnwidth]{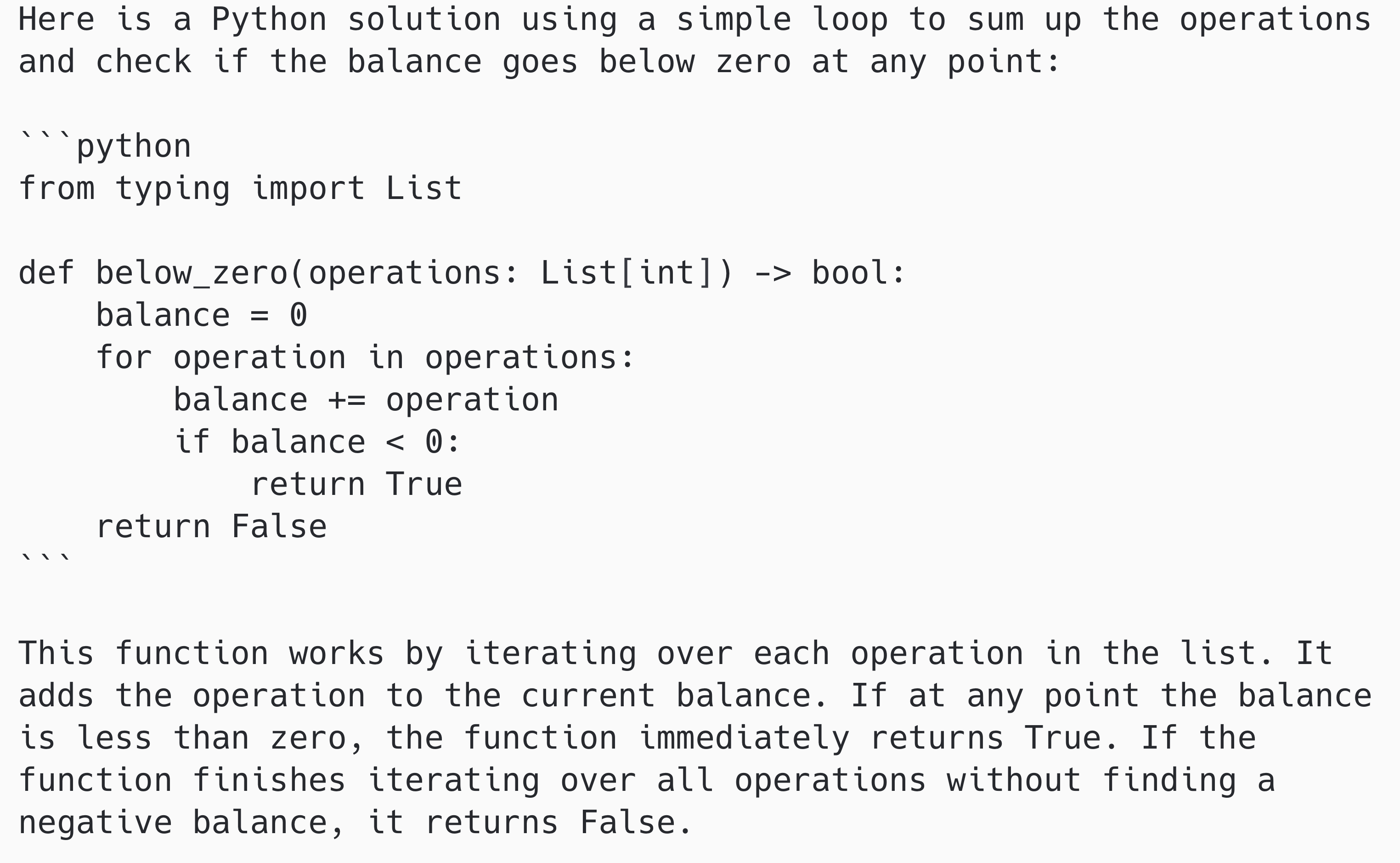}
\caption{Response generated by \deepseek-6.7B for the prompt in Fig.~\ref{fig:ex_standalone_input}.}
\label{fig:ex_standalone_output}
\end{subfigure}
\caption{{An example of independent-unit code generation for a task in the \textit{HumanEval} benchmark}}
\label{fig:ex_standalone}
\end{figure}

\begin{figure}
\centering
\begin{subfigure}{0.48\textwidth}
\centering
 \includegraphics[width=1\columnwidth]{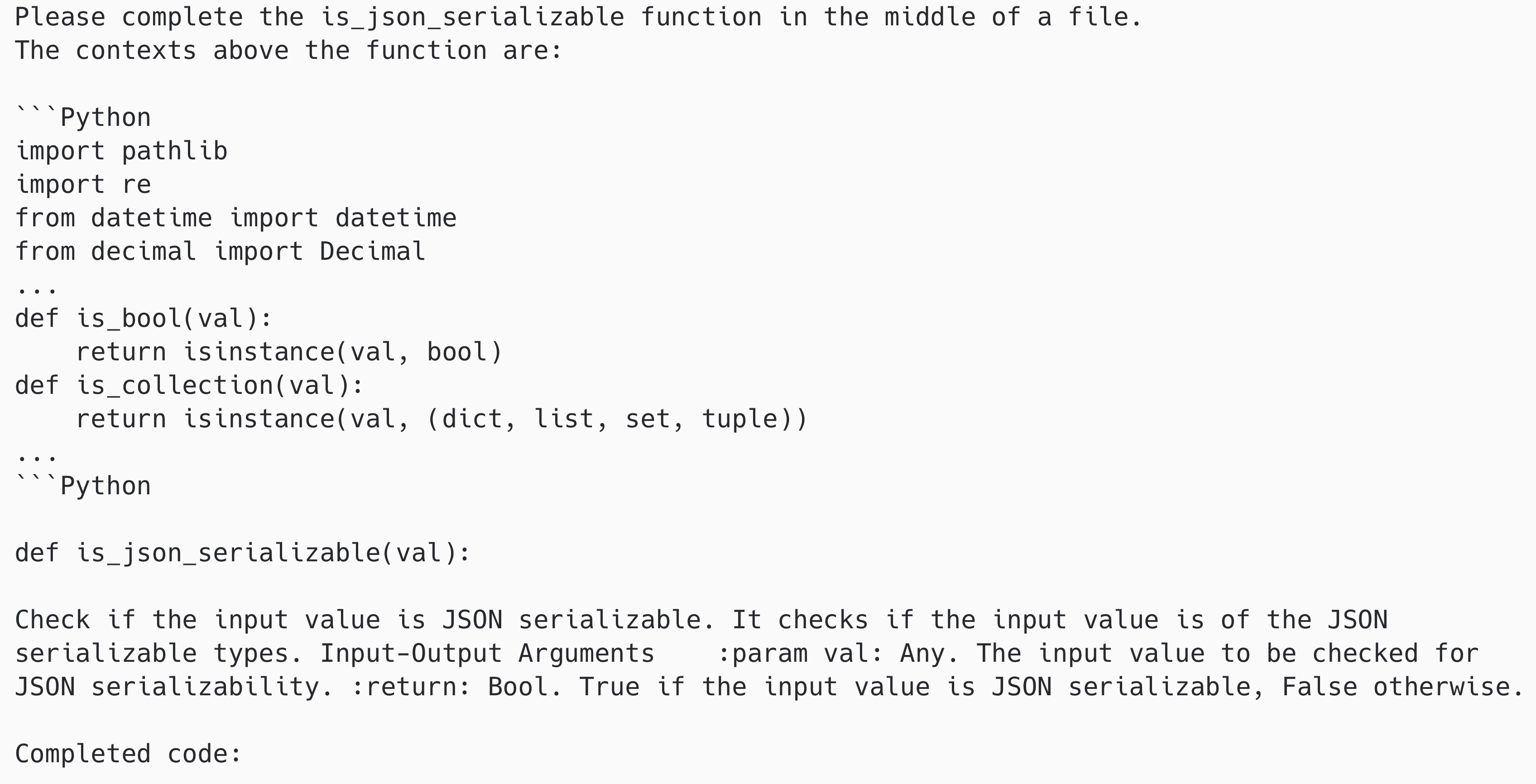}
\caption{Prompt provided to the model}
\label{fig:ex_repo_input}
\end{subfigure}\\
\begin{subfigure}{0.48\textwidth}
\centering
 \includegraphics[width=1\columnwidth]{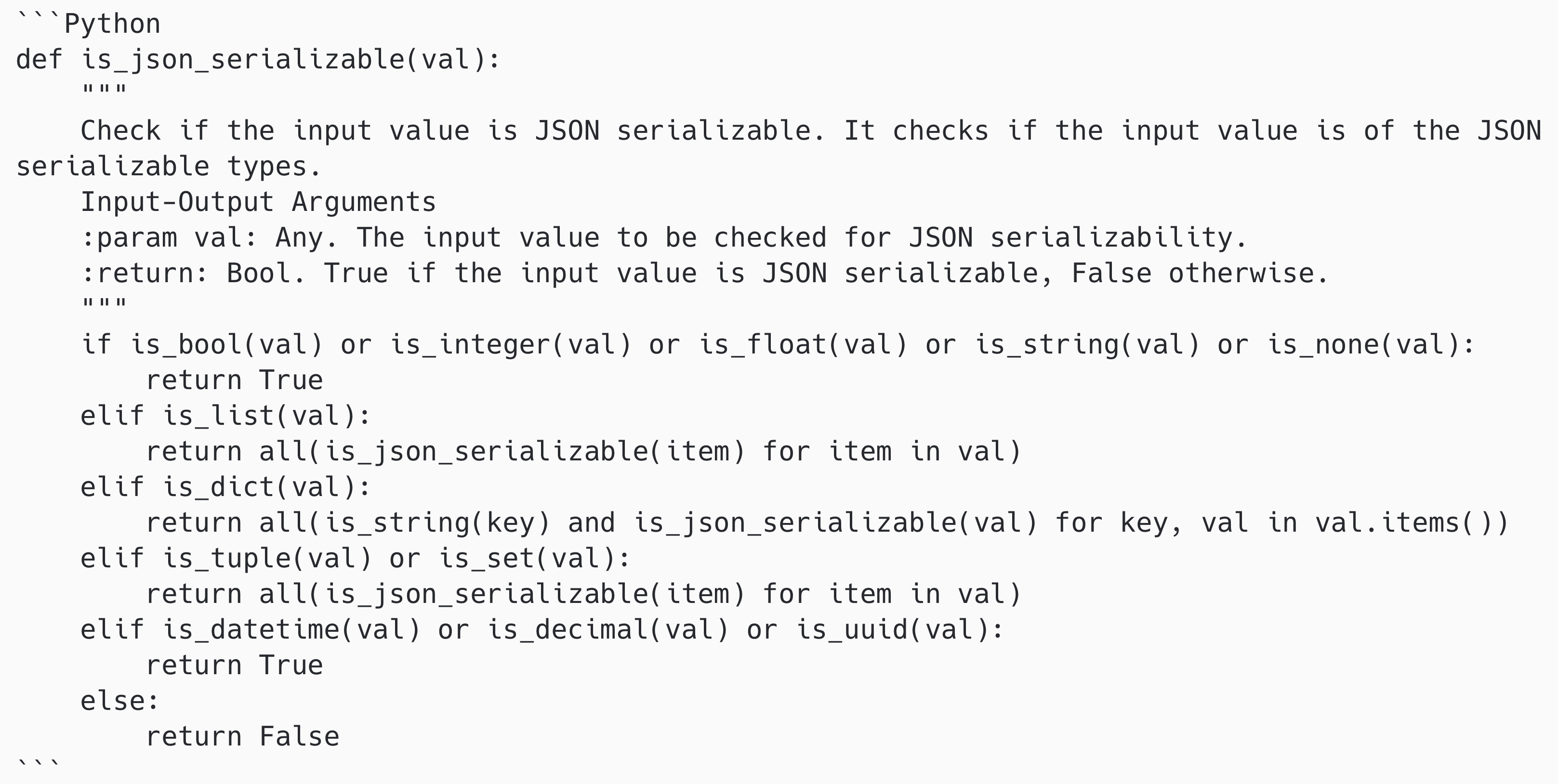}
\caption{Response generated by \deepseek-6.7B for the prompt in Fig.~\ref{fig:ex_repo_input}.}
\label{fig:ex_repo_output}
\end{subfigure}
\caption{{An example of repo-level code generation for a task in \textit{DevEval} benchmark}}
\label{fig:ex_repo}
\end{figure}

{
\mbox{Fig.~\ref{fig:ex_standalone}} and \mbox{Fig.~\ref{fig:ex_repo}} show the examples of prompts and corresponding responses generated by \mbox{\deepseek-6.7B} for the independent-unit and repository-level code generation tasks, respectively. 
Our goal is to utilize the model's internal representations produced during the generation process to predict the correctness of the generated code (\mbox{Fig.~\ref{fig:ex_standalone_output}} and \mbox{Fig.~\ref{fig:ex_repo_output}}). These internal representations refer to the hidden states associated with each generated token across different layers of the model. 
For instance, during generation, each token (e.g., keywords like \texttt{for}, \texttt{if} or identifiers like \texttt{balance} in \mbox{Fig.~\ref{fig:ex_standalone_output}}) is represented by hidden states across all 32 transformer layers of \mbox{\deepseek-6.7B}. 
Suppose \mbox{\tool} chooses the \textit{final code token}, i.e., \texttt{False} in \mbox{Fig.~\ref{fig:ex_standalone_output}}, as the focus of analysis. From this token, \mbox{\tool} extracts the corresponding hidden states at a particular layer, e.g., the last ($32^{nd}$) layer, which captures high-level semantic features that reflect the model's overall reasoning about the function. These representations are then passed to a probing classifier to predict the likelihood that the generated function is correct.
To obtain informative signals for code correctness prediction, \mbox{\tool} aims to select appropriate tokens and extract their corresponding internal representations from a suitable layer within the model.
}

\textbf{Layer Selection for Internal Representations}.
Modern (code) LLMs often contain multiple transformer layers, each contributing uniquely to the model's ability to understand and process various tasks. For example, DeepSeek-Coder-6.7B~\cite{deepseek-coder} and \codellama-7B~\cite{codellama} comprise 32 layers, while the 34B version of \codellama consists of 60 layers. 
Each layer captures different levels of abstraction and semantic representation; shallow layers often encode lower-level syntactic and lexical features, while deeper layers focus more on abstract and semantic representations.
Jin~\etal~\cite{jin2024exploring} have demonstrated the alignment of shallow and deep layers with simpler and more complex tasks, respectively. Meanwhile, Song~\etal~\cite{song2024layer} highlights the varying importance of layers within an LLM. 
These studies emphasize that not all layers contribute equally to a given task.
Thus, \textit{selecting the appropriate layer(s)} among $L$ layers of $\mathscr{M}$ for extracting internal representations is crucial for achieving optimal performance in downstream tasks.
An inappropriate choice of layer may fail to capture the necessary information, leading to suboptimal results. 
In this work, we conduct several experiments to systematically evaluate the impact of layer selection on the task of assessing the correctness of LLM-generated code (Sec.~\ref{sec:layer_selection}).

\textbf{Token Selection for Internal Representations}.
In addition to layer selection, \textit{the choice of specific tokens}
from which internal representations are extracted is another critical factor.  
Valuable signals for correctness assessment can reside in several specific tokens within the generated output or its preceding input.
Several works~\cite{yincharacterizing, yuksekgonulattention, snyder2024early} have highlighted the effectiveness of probing the hidden states of the final token in the generated answer to assess the answer's correctness. This token typically encapsulates the accumulated context and semantics of the entire generated sequence. Therefore, the hidden states of this token can provide meaningful insights for determining whether the answer is correct or not. 
Meanwhile, some other approaches~\cite{gottesman2024estimating, slobodkin2023curious} focus on the final token of the prompt, just before the response is generated. The reasoning behind this choice is that this token often reflects the model's transition from input comprehension to output generation, making its hidden states indicative of how well the model has understood the task requirement and its readiness to generate the output. 
In this work, we also systematically investigate the impact of different token selection strategies on assessing the correctness of the generated code. Our goal is to understand how token-level choices affect the effectiveness of \tool and to explore the relationship between token-specific hidden states and the correctness of the generated code (Sec.~\ref{sec:token_selection}).

\textit{In summary}, our approach, \tool, investigates how internal representations of LLMs can be effectively utilized to assess the correctness of generated code. \tool employs a representation probing classifier, which leverages these internal representations to predict whether the generated code is correct. By systematically analyzing layer selection and token selection, we aim to identify the most informative internal representations that contribute to robust and accurate correctness assessment.  Through empirical evaluations, we demonstrate how different layers and token positions impact the effectiveness of correctness assessment, providing insights into optimizing internal representation selection for improved downstream task performance.
\section{Evaluation Design}
In this research, we seek to answer the following research questions:

\noindent \textbf{RQ1. Performance Analysis:} To what extent can \tool leverage the internal representations of an LLM to assess the correctness of its generated code? How accurately does \tool compare to the baseline approaches?

\noindent \textbf{RQ2. Layer and Token Selection Analysis:} How do the representations extracted from different token positions and layer depths impact \tool's performance? 

\noindent \textbf{RQ3. Sensitivity Analysis:}  How do different factors such as programming languages and task difficulty levels impact \tool's performance?

\noindent \textbf{RQ4. Time Complexity:} What is \tool's running time?

\subsection{Dataset Construction}

\tool constructs the dataset using three popular benchmarks, \textit{HumanEval}~\cite{humaneval}, \textit{MBPP}~\cite{mbpp}, and \textit{DevEval}~\cite{li-etal-2024-deveval}. In particular, HumanEval and MBPP contain 164 and 500 standalone programming problems, respectively, while DevEval consists of 1,825 tasks of repository-level code generation.  
For each task in these benchmarks, we employed each studied model to generate 10 candidate solutions. The correctness of these generated solutions is then evaluated using the test cases provided by the respective benchmarks.
%
%
Table~\ref{tab:data_summarize} summarizes the number of correct and incorrect solutions generated by each model for the given benchmarks.
The complete dataset, including all generated code, the corresponding internal representations extracted from the studied code LLMs, and their correctness labels could be found on our website~\cite{website}.


\begin{table}\centering
\caption{{The numbers of correct (\textit{\#Cor.}) and incorrect (\textit{\#Inc.}) solutions of benchmarks}}
\label{tab:data_summarize}
\scriptsize
\begin{tabular}{l|l|r|r|r}\toprule
\textbf{Code LLM}&                                               &\textit{HumanEval}  &\textit{MBPP}    &\textit{DevEval}\\\midrule
\multirow{2}{*}{\textit{\textbf{\deepseek-1.3B}}}  &\textit{\#Cor.}      &872        &2,065   &2,092 \\
                                                  &\textit{\#Inc.}      &768        &2,701   &15,237   \\\midrule
\multirow{2}{*}{\textit{\textbf{\deepseek-6.7B}}}  &\textit{\#Cor.}      &1,294      &3,033   &4,321 \\
                                                  &\textit{\#Inc.}      &346        &1,749   &13,009 \\\midrule
\multirow{2}{*}{\textit{\textbf{\codellama-7B}}}   &\textit{\#Cor.}      &644        &1,942   &3,467 \\
                                                  &\textit{\#Inc.}      &996        &2,686   &13,863 \\\midrule
\multirow{2}{*}{\textit{\textbf{\codellama-13B}}}   &\textit{\#Cor.}      &666        &2,065   &2,849 \\
                                                  &\textit{\#Inc.}      &974        &2,855   &14,481 \\\midrule                                                
\multirow{2}{*}{\textit{\textbf{\magiccoder-7B}}}  &\textit{\#Cor.}      &1,198      &3,135   &4,113 \\
                                                  &\textit{\#Inc.}      &440        &1,752   &13,217 \\
\bottomrule
\end{tabular}
\end{table}
\subsection{{Studied Code LLMs}}


To ensure a fair, reproducible, and practical evaluation, we carefully selected LLMs specialized for code based on the following criteria. First, the models must be open-source to enable access to their internal representations during inference. Second, the models should be trained or fine-tuned on large-scale code corpora, as code-specialized LLMs are better suited for programming tasks and widely used in practice. Third, we limited the  model size to 13B parameters to accommodate hardware resource constraints.
These criteria were designed to balance model performance, accessibility, and practical usability.

Following these criteria, we selected five representative models: \deepseek-1.3B~\cite{deepseek-coder}, \deepseek-6.7B~\cite{deepseek-coder}, \codellama-7B~\cite{codellama}, \codellama-13B~\cite{codellama}, and \magiccoder-7B~\cite{magicoder}. These models are all open-source, widely adopted in both the research and industry, and have demonstrated strong performance in various code generation tasks. Each model employs advanced architectures and training methodologies designed to capture the complexities of programming languages, including syntax, semantics, and logical consistency.  For consistency across experiments, we use their instruction-tuned variants with officially released pre-trained weights from HuggingFace and do not perform any additional fine-tuning. This setup allows us to assess the models' out-of-the-box generalization capabilities and supports a controlled comparison of how internal representations relate to code correctness.
\subsection{Experimental Procedure}

\textbf{RQ1. Performance analysis:}

\textit{Baselines:} We evaluated the performance of \tool by comparing it with three baseline approaches:
\begin{itemize}
    \item {\textit{Inference-time representation-based methods}}~\cite{internal-state-2, lookback, inside, llmcheck}: {These approaches leverage the internal representations of LLMs during inference to access the correctness of generated outputs. To the best of our knowledge, no prior work has specifically explored internal representations for evaluating the correctness of code generated by LLMs. For comparative evaluation, we adopt \mbox{\llmcheck} \mbox{\cite{llmcheck}}, a method originally proposed for assessing the quality of LLM-generated outputs in natural language question answering tasks. Specifically, \mbox{\llmcheck} utilizes the diversified scoring methods, including internal attention kernel maps, hidden activations and output prediction probabilities, from different model components to maximize the capture of hallucinations.}
    
    \item \textit{Post-hoc classification-based methods}~\cite{embedding_emse22,opt-pretrained-model,pretrained-survey}: These approaches focus on capturing the semantics of generated code to predict its correctness. Specifically, pre-trained models such as CodeBERT~\cite{codebert} and CodeT5~\cite{codet5} or CodeT5+~\cite{codet5+} are employed to encode the semantics of the LLM-generated code. 
    A classifier is then trained on these embeddings to determine whether the given code is correct or not.
    \item \textit{{LLM-as-A-Judge (LLM-AJ)}}~\cite{internal-state-2}: {These approaches leverage the inherent reasoning and generalization capabilities of LLMs to access code correctness without requiring additional fine-tuning or training.
    Following the setup of prior work \mbox{\cite{internal-state-2}}, we adopt a zero-shot prompting strategy using the same prompt design, where the model is directly queried regarding the correctness of a given code snippet. The full prompt details are available on our website \mbox{\cite{website}}.
    In our experiments, we employ zero-shot inference with both the model under study (\textit{In-house LLM-AJ}) (e.g., \mbox{\deepseek}, \mbox{\codellama}, or \mbox{\magiccoder}) and an external reasoning model \mbox{\gpt} (\textit{External LLM-AJ}), which serves as an independent verifier to evaluate code correctness.}
    
\end{itemize}

For a fair comparison, we applied the same neural architecture for the classification models in both post-hoc methods and \tool. In each approach, the classifier contains the input, output layers, and two hidden layers with 128 and 64 neurons correspondingly. The training is conducted for 50 epochs with a batch size of 32 and a learning rate of $10^{-3}$.

\textit{Procedure:} 
To compare \tool's performance with the baselines, we extracted the internal state of the last generated token from the last layer of the studied LLM  for each query as a representation to assess the correctness of the corresponding generated output.
This choice is motivated by the fact that the internal state from the last layer is closest to the model's final decision, making it a logical and effective representation for evaluating the correctness of the generated code.
Moreover, the last token is generated based on both the prompt input and the entire generated output, allowing its internal state to encapsulate the most comprehensive information processed by the model. Therefore, this state is selected for conducting experiments to answer this research question.
Note that while this experiment focuses on leveraging the last token's internal state for comparison with baselines, the impacts of other tokens and layers are investigated separately in RQ2.

{
%
Note that in our comparative study, while both \mbox{\tool} and the post-hoc classification-based methods provide their code quality assessment after the code generation is complete, they are fundamentally different in terms of the input. The post-hoc classification-based methods rely exclusively on the \textit{final output code produced after the generation process}. In contrast, \mbox{\tool} gives its assessment based on the code LLMs' internal states which are \textit{the intermediate artifacts produced during the generation process}.
This design also enables \mbox{\tool} to support early code correctness assessment. For example, by analyzing the internal representations of early generated tokens/layers, correctness assessments could be made even before code generation finishes, allowing proactive interventions.}

In this experiment, we compared the performance of \tool with the baselines across multiple settings.
\begin{itemize}
    \item \textbf{\textit{Independent-unit code generation}:} This evaluates the correctness of code generated for standalone programming tasks~\cite{humaneval,mbpp}.
    \begin{itemize}
        \item \textit{Cross-benchmark}: The internal representations of a given code LLM obtained while generating code for tasks in one benchmark (i.e., MBPP) are used to train the probing classifier. The classifier is then tested on the internal representations obtained while generating code for tasks in a different benchmark (i.e., HumanEval) with the same code LLM.
        \item \textit{Cross-task}: The tasks within a benchmark are split in a 9:1 ratio, where 90\% of the tasks were used to extract internal representations for training and validating the probing classifier, and 10\% of the remaining tasks are used for testing. 
    \end{itemize}
   
    \item \textbf{\textit{Repo(sitory)-level code generation}:} This evaluates the correctness of code generated for existing projects. Unlike independent-unit code generation, repo-level tasks require LLMs to maintain contextual consistency across multiple code units, making them significantly more challenging~\cite{rambo, repocoder}. This setting measures how well \tool can handle larger, context-dependent codebases. 

\end{itemize}

\textbf{RQ2. Layer and token selection analysis:} This experiment investigates the impact of internal representations from different layers and token positions on \tool's performance in assessing code correctness. To answer this research question, we systematically evaluated the performance of \tool when it is trained and tested on internal states extracted from:
\begin{itemize}
    \item Different layers within the code  LLM, ranging from shallow to deep layers, to determine how abstraction levels influence correctness prediction. 
    
    \item Various token positions, including first and last generated token, first and last generated code token, to understand their respective contribution to accuracy.
\end{itemize}

\textbf{RQ3: Sensitivity analysis:} This experiment examines \tool's ability to generalize across programming languages. 
Specifically, we evaluated whether \tool,  trained on the internal states extracted from the generation of code in several programming languages, can effectively predict the correctness of code in a completely different language. 
In addition, we also investigated the performance of \tool with code generated for tasks of different difficulty levels.
\subsection{Metrics}

The task of accessing the correctness of code generated by LLM can be formulated as a binary classification task. Specifically, given a generated code snippet (along with related information, such as its internal representations), \tool and the baseline approaches aim to predict whether the code is correct or not. To evaluate the performance of the approaches, we adapted widely used classification metrics: \textit{Accuracy, Precision, Recall}, and \textit{F1-Score}.

Given the potential class imbalance in the dataset, as highlighted in Table~\ref{tab:data_summarize}, where the ratio of correct to incorrect code snippets may vary significantly, it is critical to ensure a fair evaluation across both labels.
To address this, we employed weighted metrics, including \textit{Weighted Accuracy}, \textit{Weighted Precision}, \textit{Weighted Recall}, and \textit{Weighted F1-score}. These metrics compare the corresponding result for each label and return the average considering the proportion for each label in the test set. 
\section{Experimental Results}
\label{sec:results}

\subsection{Performance Analysis}

\subsubsection{{Performance Comparison}}

\begin{table*}
\centering
\scriptsize
\caption{{Correctness assessment performance for \textit{\textbf{Independent-unit Code Generation}}}}
\label{tab:comparison_standalone}
\begin{tabular}{l|l|l|rrrrr}\toprule
Code LLM &\text{Setting} & &Accuracy &Precision &Recall &F1-Score \\\midrule

\multirow{10}{*}{\deepseek-1.3B} 
    &\multirow{5}{*}{\textit{Cross-benchmark}} 
        &\inhouse &0.55 &0.62 &0.55 &0.50 \\
        & &\external &\text{0.93} &\text{0.93} &\text{0.93} &\text{0.93} \\
        & &CodeBERT &0.50 &0.52 &0.50 &0.45 \\
        & &CodeT5+ &0.47 &0.52 &0.47 &0.34 \\
        & &\llmcheck &0.61 &0.63 &0.61 &0.62 \\
        & &\tool &\text{0.66} &\text{0.66} &\text{0.66} &\text{0.66} \\\cmidrule{2-7}
    &\multirow{5}{*}{\textit{Cross-task}} 
        &\inhouse &0.42 &0.42 &0.42 &0.42 \\
        & &\external &\text{0.86} &\text{0.86} &\text{0.86} &\text{0.86} \\
        & &CodeBERT &0.64 &0.64 &0.64 &0.63 \\
        & &CodeT5+ &0.66 &0.66 &0.66 &0.66 \\
        & &\llmcheck &0.58 &0.58 &0.58 &0.58 \\
        & &\tool &\text{0.73} &\text{0.73} &\text{0.73} &\text{0.73} \\\midrule

\multirow{10}{*}{\deepseek-6.7B} 
    &\multirow{5}{*}{\textit{Cross-benchmark}} 
        &\inhouse &0.32 &0.69 &0.32 &0.26 \\
        & &\external &\text{0.94} &\text{0.94} &\text{0.94} &\text{0.94} \\
        & &CodeBERT &0.35 &0.62 &0.35 &0.35 \\
        & &CodeT5+ &\text{0.66} &0.65 &\text{0.66} &0.66 \\
        & &\llmcheck &\text{0.61} &0.60 &\text{0.61} &0.61 \\
        & &\tool &0.65 &\text{0.69} &0.65 &\text{0.67} \\\cmidrule{2-7}
    &\multirow{5}{*}{\textit{Cross-task}} 
        &\inhouse &0.68 &0.78 &0.68 &0.57 \\
        & &\external &\text{0.75} &\text{0.77} &\text{0.75} &\text{0.75} \\
        & &CodeBERT &0.64 &0.60 &0.64 &0.61 \\
        & &CodeT5+ &0.69 &0.68 &0.69 &0.68 \\
        & &\llmcheck &0.65 &0.63 &0.65 &0.63 \\
        & &\tool &\text{0.79} &\text{0.79} &\text{0.79} &\text{0.79} \\\midrule

\multirow{10}{*}{\codellama-7B} 
    &\multirow{5}{*}{\textit{Cross-benchmark}} 
        &\inhouse &0.61 &0.37 &0.61 &0.46 \\
        & &\external &\text{0.97} &\text{0.97} &\text{0.97} &\text{0.97} \\
        & &CodeBERT &0.58 &0.54 &0.58 &0.54 \\
        & &CodeT5+ &0.63 &0.62 &0.63 &0.62 \\
        & &\llmcheck &0.59 &0.67 &0.59 &0.60 \\
        & &\tool &\text{0.69} &\text{0.68} &\text{0.69} &\text{0.68} \\\cmidrule{2-7}
    &\multirow{5}{*}{\textit{Cross-task}} 
        &\inhouse &0.50 &0.25 &0.50 &0.33 \\
        & &\external &\text{0.75} &\text{0.75} &\text{0.75} &\text{0.75} \\
        & &CodeBERT &0.68 &0.68 &0.68 &0.68 \\
        & &CodeT5+ &0.70 &0.70 &0.70 &0.70 \\
        & &\llmcheck &0.66 &0.45 &0.66 &0.53 \\
        & &\tool &\text{0.75} &\text{0.76} &\text{0.75} &\text{0.75} \\\midrule

\multirow{10}{*}{\codellama-13B} 
    &\multirow{5}{*}{\textit{Cross-benchmark}} 
        &\inhouse &0.41 &0.16 &0.41 &0.23 \\
        & &\external &\text{0.95} &\text{0.95} &\text{0.95} &\text{0.95} \\
        & &CodeBERT &0.61 &0.61 &0.61 &0.61 \\
        & &CodeT5+ &0.62 &0.62 &0.62 &0.62 \\
        & &\llmcheck &0.56 &0.65 &0.56 &0.56 \\
        & &\tool &0.64 &0.65 &0.64 &0.64 \\\cmidrule{2-7}
    &\multirow{5}{*}{\textit{Cross-task}} 
        &\inhouse &0.38 &0.14 &0.38 &0.21 \\
        & &\external &\text{0.75} &\text{0.75} &\text{0.75} &\text{0.75} \\
        & &CodeBERT &0.65 &0.66 &0.65 &0.65 \\
        & &CodeT5+ &0.67 &0.67 &0.67 &0.67 \\
        & &\llmcheck &0.64 &0.68 &0.64 &0.57 \\
        & &\tool &0.78 &0.79 &0.78 &0.78 \\\midrule

\multirow{10}{*}{\magiccoder-7B} 
    &\multirow{5}{*}{\textit{Cross-benchmark}} 
        &\inhouse &0.27 &0.07 &0.27 &0.11 \\
        & &\external &\text{0.94} &\text{0.95} &\text{0.94} &\text{0.94} \\
        & &CodeBERT &0.46 &0.58 &0.46 &0.49 \\
        & &CodeT5+ &0.54 &0.64 &0.54 &0.56 \\
        & &\llmcheck &0.60 &0.64 &0.60 &0.53 \\
        & &\tool &\text{0.56} &\text{0.67} &\text{0.56} &\text{0.58} \\\cmidrule{2-7}
    &\multirow{5}{*}{\textit{Cross-task}} 
        &\inhouse &0.48 &0.73 &0.48 &0.47 \\
        & &\external &\text{0.80} &\text{0.81} &\text{0.80} &\text{0.80} \\
        & &CodeBERT &0.75 &0.73 &0.75 &0.73 \\
        & &CodeT5+ &0.68 &0.68 &0.68 &0.68 \\
        & &\llmcheck &0.67 &0.68 &0.67 &0.67 \\
        & &\tool &\text{0.78} &\text{0.78} &\text{0.78} &\text{0.78} \\
\bottomrule
\end{tabular}
\end{table*}

\begin{table*}
\centering
\caption{{Correctness assessment performance for \textit{\textbf{Repo-level Code Generation}}}}
\label{tab:comparison_repo}
{%
\begin{tabular}{l|l|rrrrr}\toprule
Code LLM & &Accuracy &Precision &Recall &F1-Score \\\midrule
\multirow{5}{*}{\deepseek-1.3B} 
    &\inhouse     &0.24 &0.76 &0.24 &0.27 \\
    &\external               &\text{0.87} &\text{0.87} &\text{0.87} &\text{0.87} \\
    &CodeBERT           &0.86 &0.78 &0.86 &0.81 \\
    &CodeT5+            &0.86 &0.81 &0.86 &0.83 \\
    &\llmcheck           &0.86 &0.76 &0.86 &0.81 \\
    &\tool              &\text{0.86} &\text{0.82} &\text{0.86} &\text{0.83} \\\midrule
\multirow{5}{*}{\deepseek-6.7B} 
    &\inhouse     &0.48 &0.70 &0.48 &0.50 \\
    &\external               &0.70 &\text{0.73} &0.70 &\text{0.71} \\
    &CodeBERT           &0.71 &0.67 &0.71 &0.68 \\
    &CodeT5+            &\text{0.73} &0.69 &\text{0.73} &0.69 \\
    &\llmcheck           &0.74 &0.54 &0.74 &0.63 \\
    &\tool              &\text{0.75} &\text{0.73} &\text{0.75} &\text{0.73} \\\midrule

\multirow{5}{*}{\codellama-7B} 
    &\inhouse      &0.51 &0.65 &0.51 &0.54 \\
    &\external               &0.73 &\text{0.76} &0.73 &\text{0.74} \\
    &CodeBERT           &0.74 &0.68 &0.74 &0.69 \\
    &CodeT5+            &\text{0.76} &0.71 &\text{0.76} &0.71 \\
    &\llmcheck           &0.77 &0.59 &0.77 &0.67 \\
    &\tool              &\text{0.81} &\text{0.79} &\text{0.81} &\text{0.79} \\\midrule

\multirow{5}{*}{\codellama-13B} 
    &\inhouse      &0.19 &0.04 &0.19 &0.06 \\
    &\external               &0.77 &\text{0.79} &0.77 &\text{0.78} \\
    &CodeBERT           &0.80 &0.73 &0.80 &0.74 \\
    &CodeT5+            &\text0.78 &0.70 &0.78 &0.73 \\
    &\llmcheck           &0.80 &0.66 &0.80 &0.73 \\
    &\tool              &\text{0.81} &\text{0.77} &\text{0.81} &\text{0.78} \\\midrule

\multirow{5}{*}{\magiccoder-7B} 
    &\inhouse     &\text{0.74} &0.64 &\text{0.74} &0.65 \\
    &\external               &0.70 &\text{0.74} &0.70 &\text{0.72} \\
    &CodeBERT           &0.73 &0.68 &0.73 &0.69 \\
    &CodeT5+            &0.72 &0.67 &0.72 &0.69 \\
    &\llmcheck           &0.75 &0.56 &0.75 &0.64 \\
    &\tool              &\text{0.77} &\text{0.73} &\text{0.77} &\text{0.73} \\
\bottomrule
\end{tabular}
}
\end{table*}

Table~\ref{tab:comparison_standalone} and Table~\ref{tab:comparison_repo} show the performance of code correctness assessment for independent-unit code generation and repo-level code generation across the studied code LLMs, \deepseek-1.3B and 6.7B, \codellama 7B and 13B, and \magiccoder-7B. 
The evaluation includes LLM-as-a-judge (LLM-AJ) methods using the LLM under study (\inhouse) and a reasoning LLM \gpt (\external), post-hoc methods with CodeBERT and CodeT5+, and our approach \tool.
%
%
Overall, \textit{\tool demonstrates robust and strong performance across all metrics and evaluation settings, maintaining its top results regardless of model size or task complexity.}

%
In \textit{Cross-benchmark} (Table~\ref{tab:comparison_standalone}), although \tool lags behind \external, it significantly outperforms \inhouse and the other baseline methods.
For example, \tool achieves an Accuracy of 0.69 (for \codellama-7B), surpassing the the traditional post-hoc classification methods by a relative improvement of \textbf{10\% to 18\%}. 
%
%
In \textit{Cross-task}, \tool and \external achieve competitive performance. 
\tool obtains accuracies ranging from 0.73 (for \deepseek-1.3B) to 0.79 (for \deepseek-6.7B), significantly exceeding the performance of the post-hoc classification methods by up to \textbf{23\%}.
This superior performance highlights the effectiveness of leveraging the internal representations of the LLM, demonstrating \tool's capability to assess the correctness of generated code with high reliability.

For \textit{repo-level code generation} (Table~\ref{tab:comparison_repo}), \tool also demonstrate impressive performance, consistently surpasses the baseline approaches, including all the studied LLM-AJ methods, post-hoc methods, and \llmcheck. 
For example, \tool obtains F1-Scores ranging  from 0.73 to 0.83, while the corresponding figures of \inhouse and the post-hoc methods based on CodeBERT and CodeT5+ are between 0.27 and 0.83.
Especially for \deepseek-1.3B, \tool achieves an Accuracy of 0.86 with an F1-Score improvement of up to \textbf{ three times} over the baselines, while for \codellama-7B, \tool's Accuracy is 10\% relatively better than that of \external.
These findings underscore the advantage of leveraging internal representations, making \tool more reliable in assessing the correctness of the real-world and complex repo-level code generated by the LLMs.

In our experiments, for \inhouse with the studied LLMs (i.e., \deepseek, \codellama, or \magiccoder), we observe an imbalance in their ability to handle positive and negative cases. 
While these models can achieve competitive Precision, they struggle with Recall. 
For example, \deepseek-6.7B achieves Precision scores of 0.69 and 0.70 in the independent-unit and repo-level code generation settings, respectively, which is comparable to the other approaches. However, its Recall is up to about two times lower than the others. 
The reason is that these models tends to be over-confident in the correctness of its generated code, frequently labeling its generated code as ``correct'' even when it is actually incorrect.

In practice, LLM-AJ requires no additional training but exhibits variable performance across settings, as shown by the drop in LLM-AJ's accuracy between the independent-unit and repo-level code generation scenarios.
Meanwhile, \mbox{\tool} delivers consistently robust performance by leveraging a lightweight classifier trained on internal states. To ensure broad generalization, we evaluated \mbox{\tool} across cross-benchmark, cross-task, and repository-level scenarios, each featuring distinct data distributions. In every case, \mbox{\tool} maintained high accuracy, precision, recall, and F1-Score, demonstrating its resilience to domain shifts and varying code structures. For a practical application, practitioners can pre-train \mbox{\tool} on curated data from prior tasks or publicly available datasets. During deployment, new data can be incrementally collected and used to periodically update \mbox{\tool}, thereby improving its performance for task-specific and evolving use cases.

The traditional post-hoct approaches, which predict code correctness using the code semantics encoded by pre-trained models such as CodeBERT and CodeT5/CodeT5+, demonstrate a quite strong performance. 
Despite their strengths, these baselines are outperformed by \tool, as evidenced by consistently better performance across all models in Table~\ref{tab:comparison_standalone} and Table~\ref{tab:comparison_repo}. For repo-level code generation, \tool achieves an F1-Score of 0.73 with \deepseek-6.7B, compared to 0.69 for CodeT5+. Especially for \codellama-7B, the margin of improvement is even more significant, with \tool attaining F1-Scores of 0.79, respectively, compared to 0.71 and 0.69 for the classifier based on CodeT5+ and CodeBERT.
These results suggest that relying on pre-trained models to encode generated code, while effective to an extent, is insufficient for capturing the nuanced, dynamic, and process-oriented aspects of code correctness.

Unlike these black-box (closed-box) approaches that rely solely on final outputs, both \tool and \llmcheck are white-box (open-box) approaches that examines the LLMs' internal workings to make predictions. 
LLM-Check combines an Eigen-analysis of hidden states and attention maps with token-level uncertainty quantification to flag hallucinations without any additional training~\cite{llmcheck}. While this zero-training design makes LLM-Check immediately deployable, its performance on code-quality assessment remains lower than that of \tool up to 40\%.
%
By contrast, \tool analyzes the intermediate states of the LLM during code generation and trains a lightweight classifier on these in-process signals.
This approach uncovers hidden patterns associated with the generated code, mitigating the overconfidence often seen in LLM-AJ approach and providing a more nuanced and reliable evaluation. 
Moreover, the internal representations utilized by \tool not only encode the code features as CodeBERT and CodeT5+ but also capture the model's \textit{reasoning process}. 
This reasoning process information reflects not just dataset-specific patterns but also fundamental programming underlying programming principles. \tool can dynamically adapt to varying requirements, programming constructs, and output expectations across different tasks and benchmarks, enhancing its generality and robustness.
This adaptability enhances its generality and robustness, enabling it to consistently outperform both post-hoc classifiers and \llmcheck in our experiments across all experimental settings.
Additionally, by leveraging these \textit{in-process} signals, \tool can provide a unique advantage: \textit{it can detect potential issues and assess correctness as the code is being generated, enabling real-time quality assurance}.



{
Furthermore, \mbox{\tool} \textit{also offers a promising alternative approach for benchmarking LLMs in scenarios where ground-truth test cases are unavailable or costly to produce}. 
%
%
Indeed, in our experiments, we found that code samples labeled ``correct'' by \mbox{\tool} passed actual test cases at rates \textbf{50–250\%} higher than those labeled ``incorrect''.
Moreover, for each setting, we measured the performance of every studied code LLM in \textit{pass@1}, as well as the proportions of generated code samples labeled ``correct'' ($\mathcal{P}$) by \mbox{\tool}. We observed strong positive correlations with the Pearson correlation coefficients between \textit{pass@1} and $\mathcal{P}$ are 0.99, 0.73, and 0.97 for Cross-benchmark, Cross-task, and Repo-level settings respectively. These findings demonstrate that models ranked higher by \mbox{\tool} tend to achieve higher test-pass rates in practice.
%
%
%
As a result, developers can generate code for identical tasks from different LLMs and use \mbox{\tool} to estimate each model's relative correctness. The LLM that produces a higher proportion of predicted-correct outputs can be deemed superior. This opens up a practical and test-free solution for LLM comparison in settings where constructing tests is costly.}


\subsubsection{Code Length Analysis}

\begin{figure}
    \centering
    \includegraphics[width=\columnwidth]{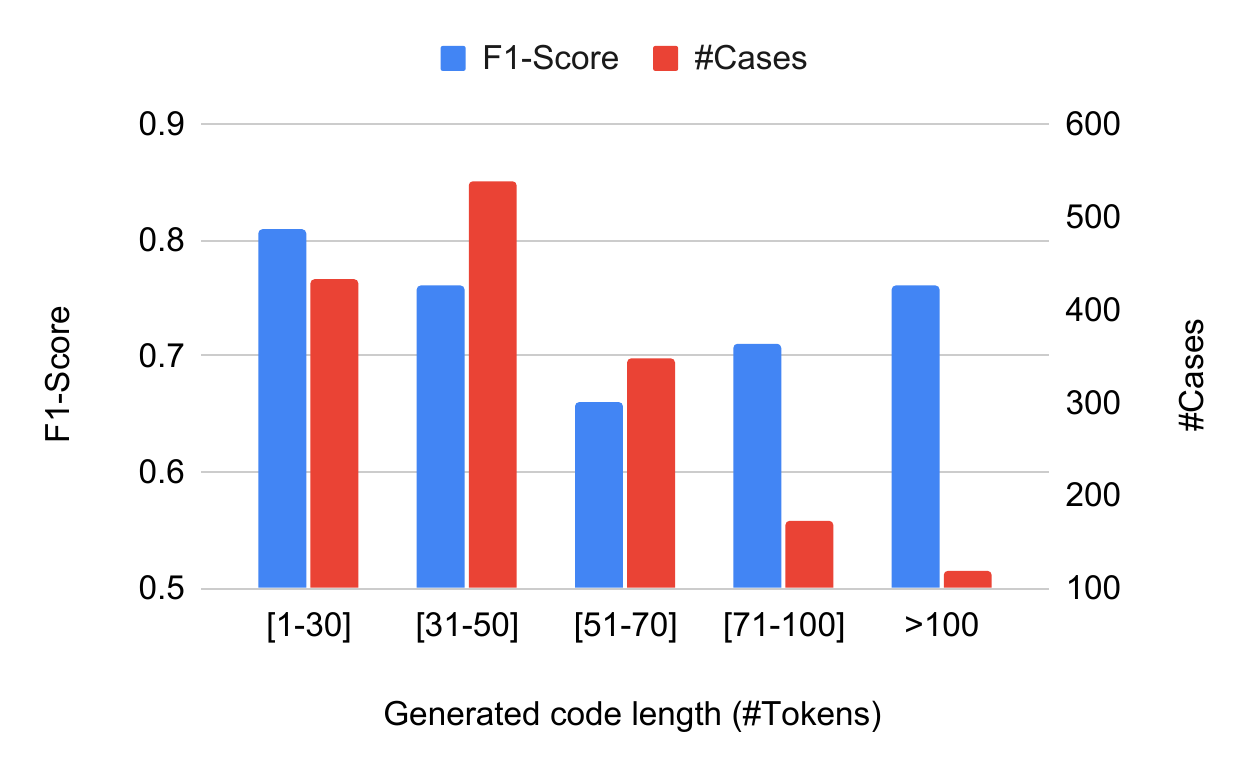}
    \caption{Impact of the length of the generated code on \tool's performance}
    \label{fig:sensitivity_code_length_impact}
\end{figure}

Figure~\ref{fig:sensitivity_code_length_impact} illustrates that \tool achieves its best performance on shorter generated code. Specifically, for the code snippets ranging from 1 to 30 tokens, \tool accurately predicts code correctness with an F1-Score of 0.81. This strong performance is attributed to the simplicity of short code snippets, which typically contain few dependencies. Thus, the semantic and syntactic features of the code can be effectively encoded within the LLM's internal states, thereby facilitating accurate predictions.
However, \tool's F1-score decreases as the code length increases, reaching its lowest point (0.66) for medium-length code (51-70 tokens). The decline in performance is likely due to the increased complexity and dependencies in longer code, which makes it more challenging for the LLM's internal representations to encode all critical information.

Interestingly, \tool exhibits performance recovery for longer code exceeding 70 tokens, despite the challenges posed by increased complexity. This improvement suggests that \tool can effectively utilize the richer contextual information and redundancy available in the internal states of longer code to reinforce key patterns and enhance correctness predictions. These results highlight the adaptability of \tool in leveraging the LLM's representations across varying lengths.

\subsubsection{Error Analysis}

\begin{table}\centering
\caption{\tool's performance on different errors}
\label{tab:error_analysis}

\begin{tabular}{l|r|rr}\toprule
\textit{Category} &\textit{F1-Score} &\#\textit{Cases} \\\midrule
Correct &0.75 &644 \\
Timed out &0.89 &10 \\
AssertionError &0.82 &825 \\
Not defined variables/attributes &0.86 &36 \\
Unsupported operands &0.82 &13 \\
Index out of range &0.73 &19 \\
Type error &0.83 &33 \\
Others &0.85 &60 \\
\bottomrule
\end{tabular}
\end{table}

{\mbox{Table~\ref{tab:error_analysis}} presents a detailed evaluation of \mbox{\tool's} performance across different error categories.
To compute F1-Score for each error type, we first categorize the generated code snippets into distinct error types according to their ground truth testing results. For each error category, we measure how accurately \mbox{\tool} identified the corresponding code snippets as incorrect.
In this table, the \textit{Correct} row represents \mbox{\tool's} performance in accurately recognizing correct code, while the remaining rows reports \mbox{\tool's} performance in detecting incorrect code in each error type.}

Overall, \tool demonstrates higher accuracy in detecting incorrect code compared to the correct one. Specifically, \tool's F1-Score for identifying correct code is 0.75, whereas its performance in detecting incorrect code reaches 0.82. This is reasonable because identifying incorrectness often requires detecting only a single error-indicating signal/pattern, making it relatively easier to classify a code snippet as incorrect. In contrast, determining correctness is inherently more challenging, as it necessitates verifying multiple aspects, including logic, structure, and functionality, to ensure the absence of errors.

Among the error types, \tool exhibits the lowest performance in detecting incorrect code caused by \textit{index out of range} errors. Specifically, its F1-Score for identifying incorrect code of this error is 0.73, which is 12\% lower than its average performance of detecting incorrect code. 
Unlike \textit{static} errors, such as undefined variables or type errors, that are inherently present in the code and can be directly captured in the LLM's internal states, index out-of-range errors are not always explicitly encoded in the internal representations of the model.
This is because such errors often depend on \textit{dynamic} factors like runtime conditions or specific input values. Therefore, effectively detecting incorrect code of this error could require a deeper understanding of program execution and data dependencies.

{
\mbox{Table~\ref{tab:error_analysis}} shows our analysis on various error categories for evaluation purposes. While \mbox{\tool} outputs a binary classification, indicating whether the generated code is correct or not, it does not provide detailed diagnostics such as error types or bug locations. As a result, in cases where the code is predicted to be incorrect, developers still need to rely on complementary analysis techniques, such as static analysis, dynamic testing, or multi-agent frameworks, to identify the root cause and guide repair.  
Despite this limitation, \mbox{\tool} offers practical benefits in several scenarios. 
First, it can be integrated into the code generation pipeline to proactively filter promising candidates and discard low-quality ones, thereby reducing the risk of presenting incorrect code to developers and minimizing debugging efforts. 
Second, \mbox{\tool} can serve as a lightweight examiner to determine whether more computationally expensive diagnostic processes are necessary.
For future work, we plan to extend \mbox{\tool} by integrating bug localization and repair phases to enhance its practical usages.
}

\subsubsection{{Potential in Improving LLM-gen. Code's Quality}}

\begin{table}
\centering
\caption{{Performance of Code LLMs when applying \textit{Correctness-Guided Generation (CG\textsuperscript{2})} using \mbox{\tool} (pass@1)}}

\label{tab:best_candidate_selection}
\resizebox{\columnwidth}{!}{%
\begin{tabular}{l|l|r|rr}\toprule
& &\textbf{Without CG\textsuperscript{2}} &\textbf{With CG\textsuperscript{2}} \\\midrule
\multirow{3}{*}{\deepseek-6.7B} &\textit{Cross-benchmark} &0.75 &0.77 \\
&\textit{Cross-task} &0.63 &0.67 \\
&\textit{Repo-level} &0.26 &0.28 \\\midrule
\multirow{3}{*}{\codellama-7B} &\textit{Cross-benchmark} &0.60 &0.61 \\
&\textit{Cross-task} &0.51 &0.57 \\
&\textit{Repo-level} &0.22 &0.24 \\\midrule
\multirow{3}{*}{\magiccoder-7B} &\textit{Cross-benchmark} &0.74 &0.76 \\
&\textit{Cross-task} &0.69 &0.73 \\
&\textit{Repo-level} &0.24 &0.27 \\
\bottomrule
\end{tabular}
}
\end{table}

To demonstrate the potential of \tool in proactively detecting and preventing bugs before they reach developers, we introduce a \textit{Correctness-Guided Code Generation (CG\textsuperscript{2})} pipeline. 
For each programming task, a LLM is employed to generate several candidates. \tool is then used to assign the correctness score to each candidate, and the highest-scoring one is selected and presented to developers. 
We compare this pipeline against the standard generation process using the \textit{pass@1} metric~\cite{humaneval}.
Note that, because \tool extracts internal states during generation and feeds them to a lightweight probing classifier with only two hidden layers, the additional time overhead is marginal (more details in Sec.~\ref{sec:time_com}).
Table~\ref{tab:best_candidate_selection} shows the performance (in pass@1) of the code LLMs, with and without CG\textsuperscript{2}, across different evaluation settings. 
%
%
%
As seen, \tool is effective in guiding the selection of high-quality code, as demonstrated by the consistent improvements of the LLMs with CG2 over the LLMs without CG2. By selecting the highest-scoring candidates predicted by \tool, the average \textit{pass@1} of the code LLMs across different settings improves by about 7\%. 
%
%

We also evaluate a filtering variant in which any solution labeled ``incorrect'' by \tool is proactively discarded before developer review. Compared to the standard code generation process, this filtering pipeline improves the average rates of passed tests by 6\%--86\%. Especially, in \textit{Repo-level} setting, the improvements are at least 53\%.

Overall, \textit{these results highlight the potential of incorporating \tool into the code generation pipeline to enhance the quality of the generated code}.

\subsection{Layer and Token Selection Analysis}

\subsubsection{Layer Selection Analysis}
\label{sec:layer_selection}

\begin{figure}
\centering
\begin{subfigure}{\columnwidth}
\centering
\includegraphics[width=1\columnwidth]{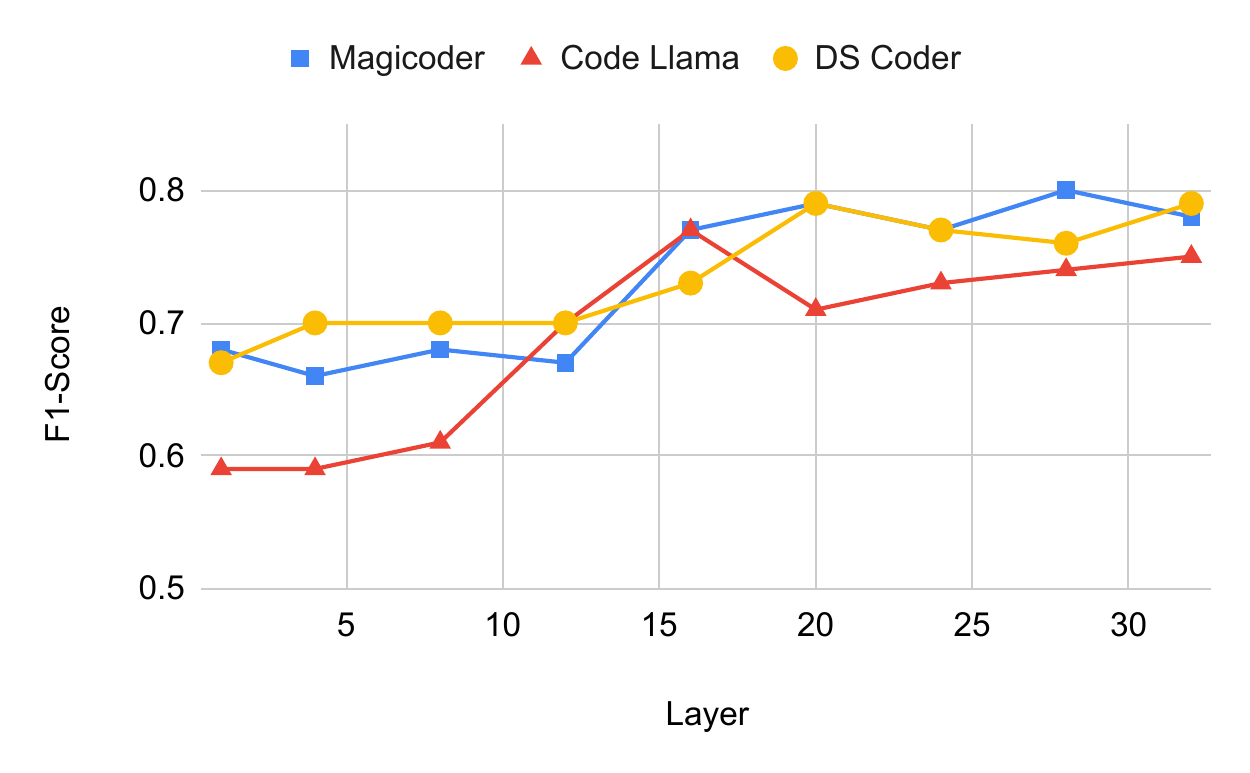}
\caption{Independent-unit code generation}
\label{fig:layer_selection_impact_standalone}
\end{subfigure}\\
\begin{subfigure}{\columnwidth}
\centering
\includegraphics[width=1\columnwidth]{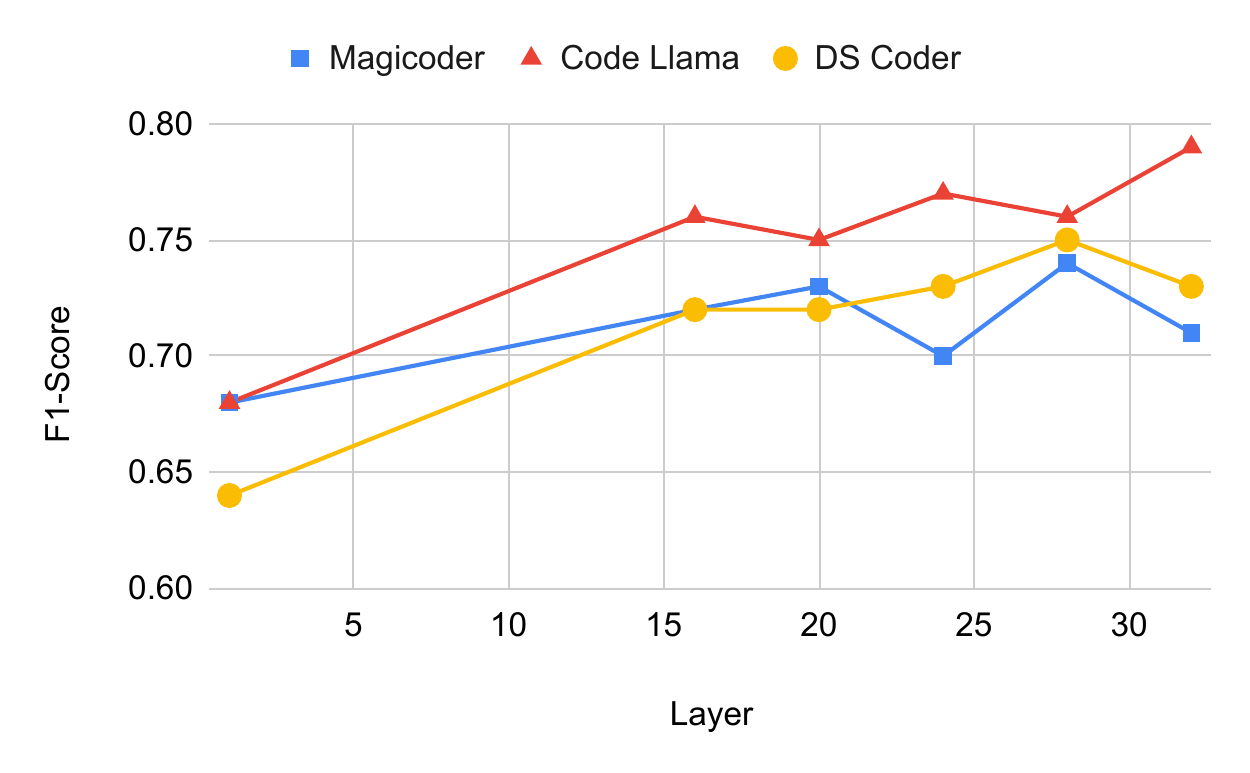}
\caption{Repo-level code generation}
\label{fig:layer_selection_impact_repo}
\end{subfigure}\\

\caption{Impact of layer selection on \tool's performance}
\label{fig:layer_selection_impact}
\end{figure}

Figure~\ref{fig:layer_selection_impact} shows the performance of \tool in correctness assessment when using hidden states from different layers of three code LLMs: \magiccoder-7B, \codellama-7B, and \deepseek-6.7B. In general, \textit{the representations from the middle layers of the models enable \tool to achieve optimal performance for both independent-unit and repo-level code generation tasks}. For independent-unit code generation (Fig.~\ref{fig:layer_selection_impact_standalone}), \tool obtains its best performance by leveraging the representations from layer 28 of \magiccoder, layer 16 of \codellama, and layer 20 of \deepseek. Similarly, for repo-level code generation (Figure~\ref{fig:layer_selection_impact_repo}), \tool attains its highest F1-Score using the representations from layer 28 for both \magiccoder and \deepseek. 
For \codellama, while \tool's peak performance is achieved with the representations from the last layer (layer 32), the middle layers (layers 16-28) still offer consistently strong results. These findings suggest that the middle layers in the LLMs typically capture a balance between the local context focused by the shallow layers and the global context encoded by the deep layers, making them optimal for assessing code correctness.

The shallow layers primarily focus on syntax and token-level details, capturing the immediate relationships between adjacent tokens (local context). While such information is useful for understanding the basic structure, it is insufficient for determining the correctness of the generated code. Consequently, leveraging the representations from the shallow layers often leads to low performance of \tool. For example, when using the representations from the first layer of \codellama, \tool achieves an F1-Score of 0.59, which is 30\% lower than its highest score.

In contrast, the deep layers often capture broader logical flow and long-range dependencies between code tokens (global context). The representations from these layers provide richer semantic information, improving \tool's ability to assess code correctness. However, the deeper the layers, the closer they are to the model's final output, making the representations at the deeper layers more task-specific. Leveraging the representations from too deep layers could make \tool to be overfitting with the training data and potentially hinder its generality. For instance, \tool observes a slight decline in performance when using the representations from the last layer of several models.

The middle layers strike a balance between the shallow and the deep layers, providing rich representations that are often suitable for downstream tasks such as code correctness assessment. However, the specific layer at which the performance of \tool peaks or dips can vary depending on the model's architecture and its pre-training objectives. This suggests that static layer selection may not be ideal, as there is no one-size-fits-all approach to selecting the best layers for \tool. Instead, \textit{\tool could benefit from a dynamic approach that adapts layer selection based on the models and task requirements}.

\subsubsection{Token Selection Analysis}
\label{sec:token_selection}

\begin{figure*}
\centering
\begin{subfigure}{1.8\columnwidth}
\centering
\includegraphics[width=1\columnwidth]{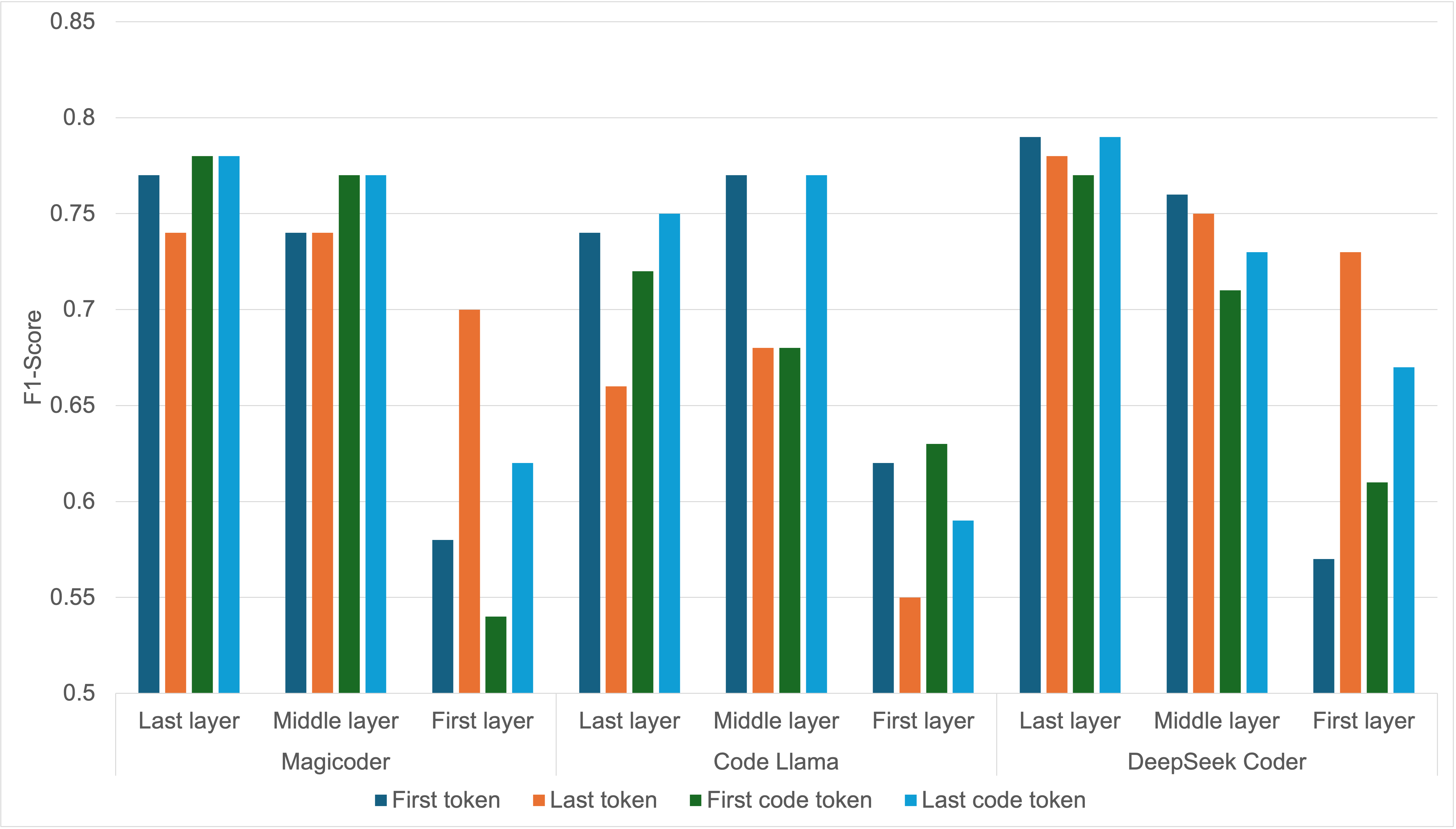}
\caption{Independent-unit code generation}
\label{fig:token_selection_impact_standalone}
\end{subfigure}\\
\begin{subfigure}{1.8\columnwidth}
\centering
\includegraphics[width=1\columnwidth]{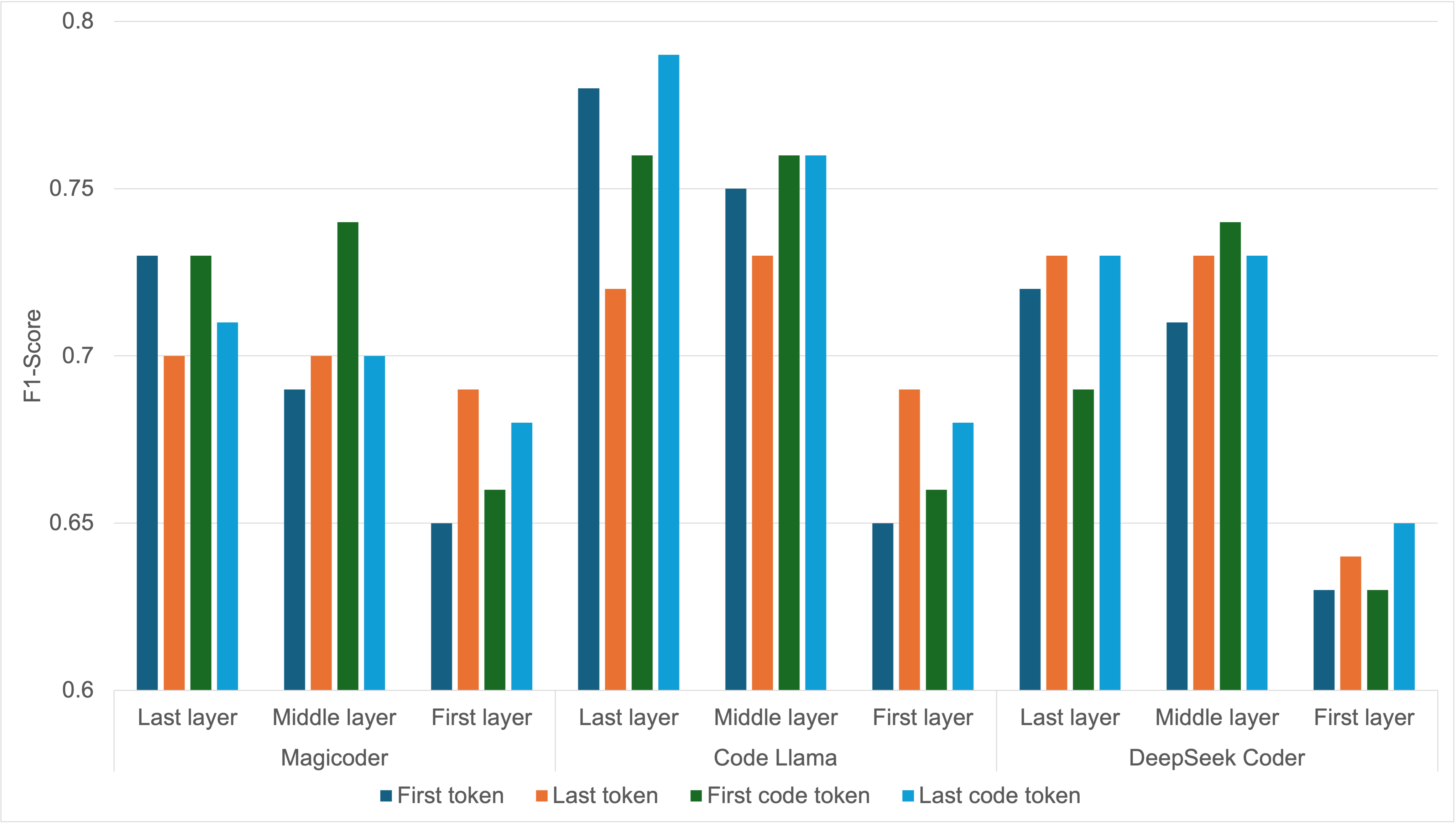}
\caption{Repo-level code generation}
\label{fig:token_selection_impact_repo}
\end{subfigure}\\

\caption{Impact of token selection on \tool's performance}
\label{fig:token_selection_impact}
\end{figure*}

Figure~\ref{fig:token_selection_impact} shows the performance of \tool when using representations of different tokens from different layers of three models, \magiccoder-7B, \codellama-7B, and \deepseek-6.7B. In all three models, the first layer corresponds to layer 1, the middle layer is layer 16, and the last layer is layer 32. 
For each layer, we extract and evaluate the performance of \tool with the representations of tokens at four specific positions: \textit{first token} and \textit{last token}, which represent the starting and ending points of the entire answer generated by the LLMs, and \textit{first code token} and \textit{last code token}, which correspond to the starting and ending points of the code segment within the LLMs' response.

In general, \textit{the deeper layer, the more stable \tool's performance is across the internal representations from different tokens; conversely, the shallower layer, the more sensitive \tool becomes to variations in token selection.}
Specifically, when using the representations from the last layers, \tool consistently achieves the highest F1-Score with the representations of the \textit{\textbf{last code token}}. Furthermore, its performance across different token representations from the last layers remains relatively stable.
In contrast, when using representations from the first layer, there is a noticeable performance gap between different tokens. For example, in assessing the correctness of code generated by \magiccoder for independent-unit tasks, (Fig.~\ref{fig:token_selection_impact_standalone}), \tool achieves an F1-Score of 0.7 with the \textit{last token} representations from the first layer, which is 30\% higher than its performance with the \textit{first code token} representations. 

Indeed, the first layer operates directly on raw token embeddings, and the interactions/relationships between tokens have not yet been captured. Due to the lack of contextualization, different tokens could convey varying levels of information, leading to different impacts on \tool's performance. This explains why \tool is highly sensitive to the token selection when using representations from the first layer of the models. Meanwhile, by the time tokens reach the last layer, their representations have been enriched with contextualized information, incorporating the relationships between tokens across the sequence. As a result, representations from the last layer are more robust and less dependent on the specific token chosen, leading to stable performance of \tool across different tokens.

However, the impact of token selection on \tool's performance varies depending on the model and layer. For instance, \magiccoder exhibits less sensitivity to token selection, while \codellama is highly affected. For example, with the representations from the last layer of \magiccoder, \tool's F1-score in predicting the code correctness for independent-unit code generation ranges from 0.74 to 0.78. However, these figures for \codellama vary more widely, 0.66--0.75. This highlights the importance of carefully selecting suitable tokens to extract representations to obtain the best performance of \tool. Similar to layer selection, \textit{token selection should be dynamically adapted to align with the specific architecture and characteristics of the model}.

\subsection{Sensitivity Analysis}

\subsubsection{Programming Language Analysis}

\begin{table}
\centering
\caption{Correctness assessment performance in F1-Score of the approaches across different languages}\label{tab:generalization_analysis}
\begin{tabular}{l|r|r|rr}\toprule
Targeted language &CodeBERT &CodeT5+ &\tool \\\midrule
CPP &0.65 &0.67 &0.82 \\
C Sharp &0.74 &0.75 &0.80 \\
Java &0.59 &0.64 &0.68 \\
JavaScript &0.66 &0.80 &0.86 \\
PHP &0.54 &0.65 &0.71 \\
Python &0.37 &0.50 &0.67 \\
Shell script &0.58 &0.58 &0.64 \\
TypeScript &0.72 &0.82 &0.88 \\
MIX &0.66 &0.67 &0.72 \\\midrule
\textbf{AVERAGE} &0.61 &0.68 &0.75 \\
\bottomrule
\end{tabular}
\end{table}
{\mbox{Table~\ref{tab:generalization_analysis}} shows the generalization performance of \mbox{\tool} and post-hoc classification-based approaches in assessing the code correctness across different programming languages. This experiment is conducted on the multi-language version of HumanEval benchmark.
Specifically, for each task in each programming language,  \mbox{\deepseek-6.7B} is employed to generate 10 candidate solutions. 
To evaluate cross-language generalization, we select one target programming language for testing, and use the generated code snippets (for post-hoc classifier) or internal states (for \mbox{\tool}) from the remaining languages for training. In addition, the row labeled ``MIX'' in the table corresponds to a setting where all generated code across all languages is combined and randomly split into training and testing sets using a 9:1 ratio. This setting enables to evaluate performance of approaches on mixed language data.}

\textit{\tool consistently outperforms the traditional post-hoc methods that rely on CodeBERT or CodeT5+ to encode the final LLM-generated code}. Across all the studied programming languages, \tool achieves an average F1-Score of 0.75, which is 12\% higher than CodeT5+ and 23\% higher than CodeBERT. Notably, for Python, \tool surpasses CodeT5+ and CodeBERT by 35\% and 81\%, respectively. These results indicate that the internal representations of the LLMs can effectively capture not only the code semantics but also the fundamental programming principles encoded in the code generation process. By levering these representations, \tool can generalize its learned knowledge to better detect code correctness across different languages.

Furthermore, all the approaches achieve their highest performance when the target language of the test set is TypeScript and their lowest performance when the target language is Python. For example, \tool's F1-Score for predicting the correctness of Python code is 0.67, which is 12\% lower than its average score. Meanwhile, for TypeScript code, \tool achieves an impressive F1-score of 0.88, exceeding its average score by 17\%. A similar trend is observed with CodeBERT and CodeT5+. This is due to Python's inherent characteristics, such as dynamic typing, indentation-based syntax, and Pythonic idioms, which significantly differ from the other languages. These features introduce ambiguity and variability, making it harder for \tool and the other approaches to generalize their learned knowledge to Python code.  
In contrast, TypeScript shares many common characteristics with other languages, such as supporting both static and dynamic typing, verbose and explicit code, and curly braces-based syntax. These shared characteristics allow \tool and the other approaches to generalize their learned knowledge more effectively, resulting in higher performance when predicting the correctness of TypeScript code.

\subsubsection{Task Difficulty}

\begin{figure}
    \centering
    \includegraphics[width=\columnwidth]{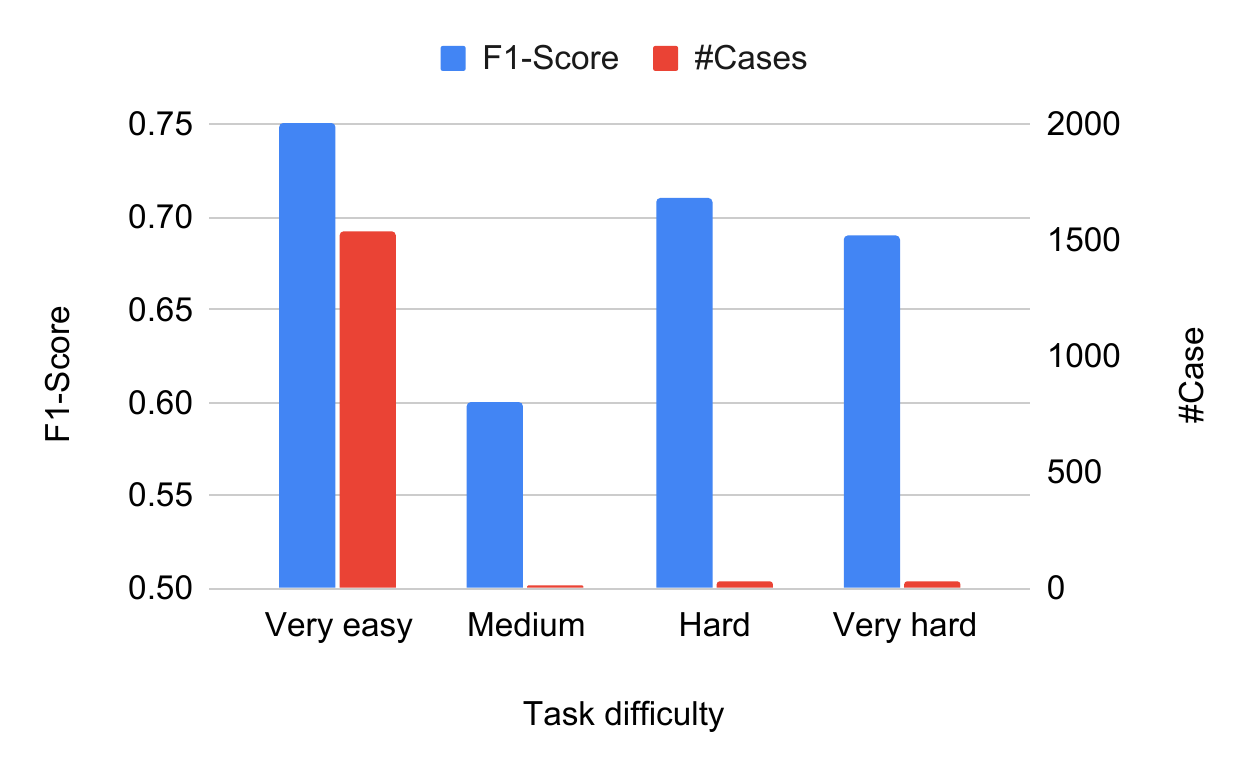}
    \caption{\tool's performance on code generated for tasks of different difficulty levels}
    \label{fig:sensitivity_task_difficulty}
\end{figure}

In this experiment, we investigate how the difficulty levels of the input tasks affect \tool's performance. To categorize task difficulty, we first prompt the studied code LLMs to classify each task in the HumanEval dataset into one of five levels: Very easy, Easy, Medium, Hard, and Very hard. The prompt used and the assigned difficulty levels for each task are available on our website~\cite{website}. We then analyze the performance of \tool in predicting the correctness of the generated code for tasks within each difficulty group.

As shown in Fig.~\ref{fig:sensitivity_task_difficulty}, \tool obtains higher performance on both easy and hard tasks, while its performance declines for medium-level tasks. Specifically, for easy tasks, \tool correctly predicts the correctness of generated code with an F1-score of 0.75, which is 25\% higher than its performance on medium-level tasks. These findings suggest that the internal representations of the code LLMs can capture its confidence about its responses. For instance, the code LLMs tend to be more confident when handling easy tasks and less confident with hard ones. This confidence signal can effectively help to assess code correctness, as reflected in \tool's high performance on these tasks. However, for medium-level tasks, the confidence signals appear to be less distinct, leading to a decrease in \tool's performance.

\subsubsection{{Prompt Variability}}
As shown in Table~\mbox{\ref{tab:comparison_standalone}} (\textit{Cross-benchmark}), differences in prompt templates between training set (MBPP) and testing set (HumanEval) could result in a decline in \mbox{\tool}'s performance, this is not observed for (output-based) post-hoc methods. For \mbox{\deepseek}-6.7B, F1-score is 0.79 under the \textit{Cross-task} setting where the prompt template remains consistent between train and test, while the corresponding figure is 0.67 when the prompt templates in the two sets are different. This is reasonable since LLMs' internal representations inherently encode prompt-specific characteristics, e.g., wording style, instruction phrasing, and contextual emphasis. Consequently, shifts in prompt format induce subtle changes in these latent signals, which in turn affect the features upon which \mbox{\tool}'s classifier depends.

To mitigate this sensitivity, we plan to augment the training set with diverse paraphrases of each prompt. This approach should stabilize \mbox{\tool}'s internal-feature space and further bolster its robustness across varied prompt formulations.
%
Preliminary experiments with mixed-prompt training show an improvement with accuracy increasing from 0.68 to 0.72, confirming that prompt augmentation can enhance \mbox{\tool}'s robustness to prompt variability.

\subsubsection{{Training Data Size}}
To understand how the amount of training data affects \tool's performance, we conducted a controlled experiment using \deepseek-6.7B in the cross-task setting. We held the test set fixed and partitioned the remaining training data into five equal folds. We then trained \tool on an incremental sequence of folds, from one fold (20\% of the data) up to all five folds (100\%).

\begin{figure}
    \centering
    \includegraphics[width=\columnwidth]{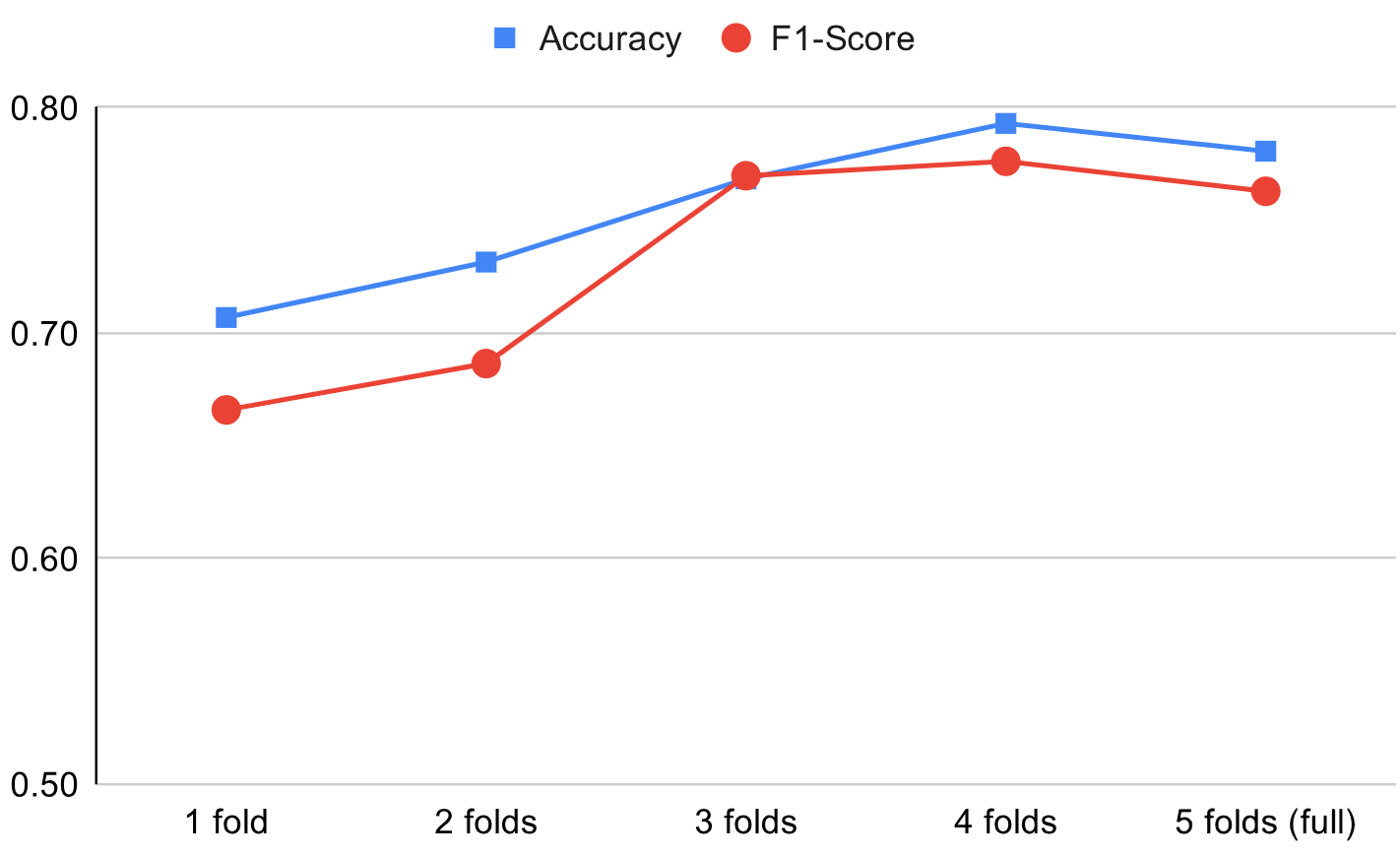}
    \caption{\tool's performance on training data size}
    \label{fig:RQ3-training-size}
\end{figure}

As shown in Fig.~\ref{fig:RQ3-training-size}, both accuracy and F1-Score rise as the training set increases from one to three folds, climbing from 0.71 (accuracy) and 0.67 (F1-score) at one fold to 0.77 for both metrics at three folds. Beyond three folds, performance gains plateau: four folds yield only marginal improvement, and the full five-fold model performs similarly.
This trend indicates that \tool quickly learns the key patterns needed to distinguish correct from incorrect code, achieving near-optimal performance with only 60\% of the available training data. Adding more data beyond that point produces diminishing returns, suggesting that \tool is data‐efficient and can be effectively trained with a relatively small corpus. In practice, practitioners can achieve robust correctness assessment without requiring large labeled datasets, reducing annotation cost and accelerating deployment in resource-constrained settings.

\subsection{Time Complexity}
\label{sec:time_com}

The time complexity of \tool is primarily determined by three factors: preparing training data and extracting internal representations from the studied code LLMs, training the probing classifier, and performing inference.

Preparing the training data for \tool and the studied post-hoc approaches on an NVIDIA A6000 GPU with 48GB of VRAM took approximately 86 hours for the entire dataset of 24,890 code units generated by the LLMs. The extraction of internal representations from the studied code LLMs was performed during code generation without affecting the process of the code LLMs.
%
%
All subsequent experiments, including probing classifier training and inference, were conducted on a single NVIDIA P100 GPU with 16GB of VRAM.
\tool demonstrates high efficiency in both training and inference due to its lightweight probing classifier, which consists of two hidden layers (128 and 64 neurons). The training process takes approximately 1 minute, as there is no embedding computation overhead. In contrast, classifiers built on pre-trained models, such as CodeBERT and CodeT5+, require significantly more time, taking about 83 minutes to train, and the embedding time took about 98\% of this amount of time.
\tool's inference process is extremely fast, requiring only 0.6 milliseconds per generated code unit. This highlights the feasibility of \tool for real-time, in-process assessment of code correctness during LLM generation, making it well-suited for practical deployment scenarios such as in interactive and real-time coding environments.


\subsection{Threats to Validity}

The main threats to the validity of our work consist of internal, construct, and external threats.

\textbf{Internal Validity}. Threats to internal validity include the hyperparameter selection for both \tool and the baselines. To mitigate this threat, we systematically explored various hyperparameter configurations for \tool and evaluated its performance under varied conditions (as detailed in Sec.~\ref{sec:results}). For baseline methods, we reused implementations and settings from the original papers, such as those for CodeBERT and CodeT5+~\cite{codebert, codet5+}, ensuring fair comparison.
Additionally, potential threats could arise from the procedures used to extract the internal representations of LLMs and their correlation with code correctness. To address this, we utilized the widely used Pytorch libraries to extract and interpret the intermediate states of LLMs during code generation. These methods were validated on diverse benchmarks and cross-verified using established practices in the field.
To further minimize the risk of errors in implementation, our codebase has undergone extensive internal review, and we have made it publicly available~\cite{website}. 

\textbf{Construct Validity}. Threats to construct validity relate to the appropriateness of our evaluation metrics and procedures. We employed widely recognized metrics, including \textit{Accuracy}, \textit{Precision}, \textit{Recall}, and \textit{F1-Score}, to measure the performance of the approaches, as these are standard benchmarks in evaluating classification models. 
Additionally, we designed experiments to include both standalone and repository-level code generation tasks, ensuring comprehensive evaluations.
However, our evaluation primarily relies on automated metrics and does not yet incorporate human assessments. While automated metrics provide objectivity and scalability, human evaluation could offer deeper insights into the practical utility of \tool in real-world coding scenarios. We plan to include such assessments in future work, engaging experienced developers to evaluate the generated code based on factors like functionality, readability, and adherence to programming standards.
Also, a potential threat is our procedure labeling the correctness of code generated by the studied code LLMs, particularly the ``correct/passed'' label, due to the potential low-quality tests. To mitigate this threat, we utilized the widely-used benchmarks~\cite{humaneval,mbpp,li-etal-2024-deveval} with adequate test sets.
Another potential limitation is that our controlled experimental setup may not fully reflect real-world software development environments. To address this, we evaluated \tool on multiple representative benchmarks, covering diverse programming constructs and code generation requirements.

\textbf{External Validity}. Threats to external validity concern the generalizability of our results to different LLMs and application domains. To mitigate this, we selected three widely used and representative open-source code LLMs: \deepseek, \codellama, and \magiccoder~\cite{deepseek-coder,codellama,magicoder}. These models have demonstrated strong performance across diverse code-related tasks, enhancing the applicability of our findings within the code generation domain.
A further potential threat arises from our focus primarily on models with parameter sizes up to 7B, due to hardware constraints. While this ensures feasibility within our experimental setup, it may limit the applicability of our conclusions to larger-scale models. As part of future work, we plan to investigate the scalability of \tool to larger LLMs and expand our experiments to a broader range of programming languages and codebases to enhance the generalizability of our approach.
{One potential threat to the robustness of \mbox{\tool} 
lies in the variability of input prompts. 
However, this threat can be partially mitigated by our evaluation across multiple benchmarks, i.e., HumanEval, MBPP, and DevEval, which exhibits diverse prompt styles and task descriptions. The consistent performance of \mbox{\tool} across these diverse benchmarks indicates a certain degree of robustness to prompt variation. Nonetheless, it remains important to systematically evaluate prompt sensitivity and we plan to investigate its impact on \mbox{\tool}'s performance in future work.}

\section{Related Work}
\textbf{LLM-based Code Generation}.
Code generation is an essential task in software development, aimed at assisting programmers by suggesting subsequent code units based on the current context~\cite{naturalness,bigcode,ase19,arist,tu2014localness}. 
Recently, \textit{Large Language Models for code (Code LLMs)}~\cite{codellama,deepseek-coder,code-gen-app,code-llm-survey} such as Codedex, \codellama, and \deepseek have emerged as promising tools for code generation/completion, offering the potential to automate repetitive tasks and increase developer productivity. These models, trained on massive datasets and equipped with billions of parameters, have demonstrated remarkable performance in code generation and completion~\cite{zhang2023repocoder, jiang2023impact, schafer2023empirical}.
Code LLMs have been deployed as auto-completion plugins, such as Github Copilot and CodeGeeX2~\cite{codegeex} in modern IDEs, and successfully streamline real-world software development activities to a certain degree~\cite{llm-code-quality, reacc,tang2023domain,bugs-llm-gen-code}.
Beyond standard code generation and completion, repo(sitory)-level code generation adds another layer of complexity as it requires integrating repository-specific elements such as user-defined APIs, inter-module dependencies, and project-specific code conventions~\cite{repocoder,tang2023domain,repoformer,reacc}.
To address the challenges of repo-level code completion, recent approaches have applied the Retrieval-Augmented-Generation (RAG) strategy.
By leveraging repository-specific knowledge, RAG-based methods enhance the capabilities of Code LLMs, significantly improving their performance in completing repo-level code tasks~\cite{repocoder, rambo, repoformer,repohyper,rlcoder}.

\textbf{Quality Assurance for LLM-generated Code}. The quality of code generated by LLMs has garnered significant attention due to the increasing reliance on AI-generated code in real-world applications. Several empirical studies have investigated bugs and security issues in LLM-generated code, highlighting the potential risks associated with their use~\cite{lost-at-c, empirical-study-2, llm-gen-code-emp-study, calibration, vulnerabilities-copilot}. These studies provide a detailed analysis of common flaws and vulnerabilities, emphasizing the need for robust evaluation and mitigation strategies.

Hallucinations, where LLMs produce plausible yet incorrect or non-functional code, are another prominent issue in code generation~\cite{hallucination-in-code}. Understanding and addressing these hallucinations is essential for ensuring the reliability of LLMs in software development. Furthermore, research has shown that AI-generated code often contains vulnerabilities, necessitating a deeper investigation into its safety implications~\cite{ai-gen-code-safety}. For instance, comparative studies have assessed the security vulnerabilities of ChatGPT-generated code against traditional sources like StackOverflow, revealing notable differences in the safety and correctness of the generated code~\cite{chatgpt-vs-stackoverflow-vul}.
Efforts have also been made to analyze the cognitive processes of LLMs during code generation. A study explored whether LLMs pay attention to code structures in a manner similar to human programmers, providing insights into their decision-making and error tendencies~\cite{llm-attention}. Uncertainty analysis for LLMs has been proposed to measure their confidence in generated outputs, which could aid in identifying potentially unreliable code~\cite{uncertainty-measure}.

To enhance the reliability of LLM-generated code, researchers have introduced methods like ``slow-thinking,'' which instructs LLMs to perform step-by-step analyses of code functionality to detect errors more effectively. Multi-agent frameworks have also been developed to secure code generation using static analysis and fuzz testing, such as AutoSafeCoder~\cite{autosafecoder}. Similarly, LLMSecGuard combines static code analyzers with LLMs to provide enhanced code security, leveraging the strengths of both approaches to mitigate vulnerabilities and ensure code robustness~\cite{llm-security-guard}.

\textbf{Bug/Vulnerability Detection in Code}. Detecting bugs and vulnerabilities in code is a critical aspect of software quality assurance, aiming to identify potential issues before they lead to security breaches or system failures. Various methods have been proposed to assess the vulnerability of code components, such as files, functions, methods, or even individual lines of code~\cite{survey_papers, poster, vuldeeppeaker, vulsniper, devign, are_we_there, mvd, issta_22, type-ist, codejit, vultype, oppsla19,COSTA}. Devign~\cite{devign} employs a graph-based approach leveraging the Code Property Graph (CPG) to predict vulnerabilities effectively. Similarly, tools like VulDeePecker~\cite{vuldeeppeaker} and SySeVR~\cite{sysevr} focus on detecting vulnerabilities at the slice level, enabling more fine-grained identification.
To enhance the granularity of vulnerability detection, graph-based models have also been developed. IVDetect~\cite{ivdetect} applies a graph-based neural network to detect vulnerabilities at the function level while using interpretable models to pinpoint vulnerable statements within suspicious functions. Similarly, VELVET~\cite{velvet} adopts graph-based techniques to detect vulnerabilities directly at the statement level.

More recent approaches integrate pre-trained models like CodeBERT~\cite{codebert} and CodeT5~\cite{codet5}, which have demonstrated significant effectiveness in both function-level and statement-level vulnerability detection tasks. For example, LineVul~\cite{linevul} and LineVD~\cite{linevd} leverage CodeBERT to encode code representations, outperforming traditional methods like IVDetect in identifying vulnerabilities at finer levels of granularity. The success of these pre-trained models is largely attributed to their ability to capture rich contextual information and encode both syntactic and semantic characteristics of code efficiently. As a result, models like CodeBERT and CodeT5 have established themselves as powerful baselines for advancing the field of bug and vulnerability detection, driving progress toward more accurate and effective software quality assurance tools.

\textbf{Analyzing LLM Internal States in Hallucination Detection and Mitigation}. %
Hallucinations, where LLMs generate plausible but inaccurate or false information, remain a critical challenge in the deployment of these models for real-world applications~\cite{llm-evaluation, llm-evaluation2}. 
TrustLLM~\cite{TrustLLM} provides a comprehensive study on trustworthiness in LLMs, covering multiple dimensions such as trust principles, benchmarks, and evaluation methods for mainstream LLMs. 
Recently, several studies have introduced white-box/open-box approaches in detecting and mitigating hallucination of general-purpose LLMs, which require analyzing the internal states of LLMs to uncover the root causes and enhance their reliability~\cite{inner-working, probing}. Azaria~\etal~\cite{internal-state} introduce a classifier trained on the hidden layer activations of LLMs that can predict the probability of a statement being truthful. 
Another study by Ji \etal~\cite{internal-state-2} explores the internal mechanisms across diverse Natural Language Generation (NLG) tasks, analyzing over 700 datasets to uncover patterns in LLM behavior.
INSIDE~\cite{inside} leverages the dense semantic information within LLMs' internal states to detect hallucinations. FactoScope~\cite{factoscope} further investigates factual discernment by measuring the inner states of LLMs, providing insights into their capacity to differentiate truth from hallucination.

{While our work and prior studies\mbox{~\cite{internal-state-2, inside, internal-state, kim2024detecting}} similarly leverage internal representations of LLMs to detect hallucination, \mbox{\tool} fundamentally differs in both \textit{task objective} and \textit{methodological approach}. These exiting studies~\mbox{\cite{internal-state-2, inside, internal-state, kim2024detecting}} primarily focus on hallucination detection in NLG, where outputs are typically unstructured and correctness is subjective, relying on human judgment or alignment with external factual knowledge. In contrast, \mbox{\tool} targets code generation, in which correctness is objectively defined through syntax, logic, and runtime behavior, as can be verified by functional test cases. 
Moreover, due to the nature of NLG tasks, prior works mainly investigate semantic consistency or factual alignment within model responses. Azaria \mbox{\etal~\cite{internal-state}} detect hallucinations by training a classifier on the internal representations of both true and false factual statements. INSIDE \mbox{~\cite{inside}} measures the semantic diversity and consistency across multiple responses to the same question to assess the risk of hallucination. 
However, these techniques are not directly applicable to code. Unlike natural language, where factual inconsistency often indicates hallucination, a programming task may have multiple correct solutions that differ in implementation details or algorithms. As a result, inconsistency across generated responses does not necessarily imply incorrectness. This key difference highlights the need for a specific approach for code generation assessment like \mbox{\tool}.}

Beyond analyzing internal states, alternative methods have been proposed to enhance the truthfulness and reliability of LLM outputs. LLMGuardrail~\cite{LLMGuardrail} integrates causal analysis and adversarial learning to develop unbiased steering representations for improving output quality. LookBack~\cite{lookback} uses attention maps to identify and mitigate contextual hallucinations. Fact-checking through token-level uncertainty quantification~\cite{fact-checking-uncertainty} provides a complementary approach for verifying the factual accuracy of LLM-generated outputs. 
Inference-Time Intervention (ITI)~\cite{ITI} focuses on enhancing LLM truthfulness by dynamically modifying model activations during inference.

{In general, these white-box approaches~\mbox{\cite{lookback, inside, internal-state, internal-state-2}}, leverage internal neural signals of LLMs such as hidden layers, attention weights, or token embeddings, etc. to explore the \textit{reasoning process} of LLMs during inference. This reflects a model-centric perspective, which seeks to understand how the model internally ``thinks'' and makes decisions, even when its reasoning is not explicitly expressed in the output. 
Meanwhile, chain-of-thought (CoT) prompting is an output-centric technique that encourage the model to produce step-by-step reasoning as part of its response~\mbox{\cite{cot}}. CoT treats this verbalized reasoning as a proxy for the model's internal thought process. While both approaches aim to uncover how LLMs arrive at their outputs, they differ in whether reasoning is inferred from internal representations or observed directly in the generated response.}

\section{Conclusion}
This work introduces \tool, a novel framework that leverages the internal representations of code LLMs to assess the correctness of generated code. Unlike traditional black-box approaches, \tool adopts a white-box methodology, systematically analyzing the intermediate states of specialized code LLMs such as \deepseek, \codellama, and \magiccoder. Our empirical findings demonstrate that these internal signals capture latent information, including fundamental programming principles, which strongly correlate with the reliability of the generated code.
\tool not only provides a more nuanced and reliable evaluation of code correctness but also exhibits strong adaptability and robustness across diverse benchmarks and tasks. By unlocking the potential of in-process signals, \tool bridges the gap between code generation and quality assurance, offering a proactive approach to enhancing the reliability of LLM-generated code. Our framework sets the foundation for future research into leveraging the latent capabilities of LLMs to improve AI-driven software development.
\section*{Acknowledgement}
This research is supported by Vietnam National Foundation for Science and Technology Development (NAFOSTED) under grant number 102.03-2023.14.
%
%
This research is also partly supported by OpenAI's Researcher Access Program.

\bibliographystyle{elsarticle-num}

\bibliography{references,ref-son}

\end{document}